\documentclass[aps,prl,10pt,twocolumn,showpacs,preprintnumbers,amsmath,amssymb,superscriptaddress]{revtex4-1}%
\usepackage{lineno}
\usepackage{graphicx}  
\usepackage{dcolumn}   
\usepackage{bm}        
\usepackage{amssymb}   
\usepackage{amsmath}
\usepackage{blkarray, multirow, graphicx, diagbox, color, xcolor, colortbl}
\usepackage[caption=false]{subfig}
\usepackage{bbm, bbold}
\usepackage{ifthen}
\usepackage[colorlinks, linkcolor = blue, citecolor = blue, filecolor = black, urlcolor = blue]{hyperref}
\usepackage{xkeyval}
\usepackage{moreverb}
\usepackage{rotating}
\usepackage{slashbox}
\usepackage{xspace}
\usepackage{nicefrac}
\usepackage[]{units}
\usepackage{physics}
\usepackage{braket}
\usepackage[inline]{enumitem}
\usepackage{tabto}
\usepackage{listings}
\usepackage{xstring}
\usepackage{glossaries}
\usepackage{hyphenat}

\glsdisablehyper

\def\ReplaceStr#1{%
	\IfSubStr{#1}{p}{%
		\StrSubstitute{#1}{p}{.}}{#1}}

\usepackage[customcolors]{hf-tikz} 
\usepackage{tikz}
\usepackage{calc}
\usetikzlibrary{external}
\graphicspath{{figures/autogen/}}
\usepackage{pgffor}
\usepackage{pgfplots}
\pgfplotsset{compat=newest}
\usepackage{pgfplotstable}
\usepgfplotslibrary{groupplots}
\usepgfplotslibrary{fillbetween}

\tikzstyle{n} = [draw,shape=ellipse,minimum size=1.5em,inner sep=0pt,fill=white!20, minimum width=2.5em]
\tikzstyle{Init} = [n,color=green,fill=green!20,text=black]
\tikzstyle{Fin} = [n,color=red,fill=red!20,text=black]
\tikzstyle{Ghost} = [minimum size=1.5em,inner sep=0pt,color=white,text=black]
\tikzstyle{Multiple} = [draw,shape=rect,minimum size=2em,inner sep=0pt]

\tikzstyle{ghostA} = [text=red!70,thick, minimum size=2*(5pt-\pgflinewidth), inner sep=0pt, outer sep=0pt]
\tikzstyle{ghostB} = [text=blue!70,thick, minimum size=2*(5pt-\pgflinewidth), inner sep=0pt, outer sep=0pt]
\tikzstyle{siteA} = [draw=red!70,circle,thick, minimum size=2*(5pt-\pgflinewidth), inner sep=0pt, outer sep=0pt]
\tikzstyle{siteB} = [draw=blue!70,circle,thick, minimum size=2*(5pt-\pgflinewidth), inner sep=0pt, outer sep=0pt]
\tikzstyle{operatorA} = [cross out, draw=red!70, thick, minimum size=2*(5pt-\pgflinewidth), inner sep=0pt, outer sep=0pt]
\tikzstyle{operatorB} = [cross out, draw=blue!70, thick, minimum size=2*(5pt-\pgflinewidth), inner sep=0pt, outer sep=0pt]

\tikzstyle{site} = [circle,thick,inner sep=0.2pt,minimum width=1.25em,font=\footnotesize,draw=blue!50!white,fill=blue!15!white,text opacity=1]
\tikzstyle{unsite} = [circle, outer sep=0pt,inner sep=0.2pt,minimum width=1.25em]
\tikzstyle{ghost} = []
\tikzstyle{op} = [regular polygon, regular polygon sides=4, draw=orange!50, fill=orange!20, thick, inner sep=0.2pt, minimum width=1.25em, minimum height=1.5em,font=\footnotesize]
\tikzstyle{ld} = [inner sep=1pt, font=\small]
\tikzstyle{intersite} = [regular polygon, regular polygon sides=4, shape border rotate= 45, draw=black!50,fill=black!20,thick,inner sep=0pt,minimum width=1.5em]

\definecolor{colorA}{rgb} {0.48,0,0.5275}
\definecolor{colorB}{rgb} {0.11,0.663,0.51}
\definecolor{colorC}{rgb} {0.3373,0.7059,0.9137}
\definecolor{colorD}{rgb} {0.902,0.8735,0.1}
\definecolor{colorE}{rgb} {0.9451,0.902,0.3255}
\definecolor{colorF}{rgb} {0.3373,0.3255,0.902}
\definecolor{colorG}{rgb} {0.9451,0.3255,0.3373}
\definecolor{colorH}{rgb} {0.11,0.3255,0.3373}

\def\gpmarkers{{"+","x","star","square","square*","o","*","triangle","triangle*"}}
\def\gpcolors{{"colorA","colorB","colorC","colorD","colorE"}}

\usetikzlibrary
{
	calc,
	decorations,
	pgfplots.patchplots,
	plotmarks,
	patterns,
	positioning,
	petri,
	arrows,
	intersections,
	decorations.markings,
	backgrounds,
	fit,
	matrix,
	graphs,
	shapes.geometric,
	decorations.pathreplacing, 
	decorations.pathmorphing,
	shapes.misc,
	shapes.multipart,
	shapes,
	through,
	tikzmark
}

\pgfplotsset{
        cycle from colormap manual style/.style={
            x=3cm,y=10pt,ytick=\empty,
            stack plots=y,
            every axis plot/.style={line width=2pt},
        },
}

\tikzset{->-/.style={decoration={
			markings,
			mark=at position .5 with {\arrow{>}}},postaction={decorate}}}

\tikzset{-<-/.style={decoration={
			markings,
			mark=at position .5 with {\arrow{>}}},postaction={decorate}}}

\tikzstyle{orientedsnake} = [
decorate, 
decoration={snake},
->
]  
\tikzstyle{orientedshortarrow} = [
decoration={markings,
	mark=at position .33 with {\arrow{>}}},
postaction={decorate}
]  
\tikzstyle{orientedlongarrow} = [
decoration={markings,
	mark=at position .67 with {\arrow{>}}},
postaction={decorate}
]
\tikzset{dbl/.style={double,
		double equal sign distance,
		-implies,
		shorten >=10pt,
		shorten <=10pt}}
\tikzset{
	between/.style args={#1 and #2}{
		at = ($(#1)!0.5!(#2)$)
	}
}
\pgfmathdeclarefunction{linearFct}{2}%
{%
	\pgfmathparse{#1*x+#2}%
}
\pgfmathdeclarefunction{logFct}{2}%
{%
	\pgfmathparse{#1*log10(x)+#2}%
}
\pgfmathdeclarefunction{pointmetalog}{3}%
{%
	#1
}%

\newcommand{\nodagger}[0]{{\phantom{\dagger}}}
\newcommand{\noprime}[0]{{\phantom{\prime}}}

\newcommand{\pprime}[0]{{\prime\prime}}

\newcommand{\fv}{\vphantom{\braket{\tilde n_j, b(\tilde n_j)}}\hspace{-0.2em}}
\newcommand{\fvN}{\vphantom{\braket{\omega| \hat N_\odot| \omega }_\odot}\hspace{-0.2em}}
\newcommand{\fvR}{\vphantom{\braket{\omega| \hat R_{\odot} | \omega }_\odot}\hspace{-0.2em}}

\pgfmathdeclarefunction{peierlspotential}{6}%
{
	\pgfmathparse
	{
		#2 / (#3 * #5) 
		* sin(deg(#1 * #3 - #3 * #5 * (x)))
		* exp(-( ( #1 - #5 * ((x) - #4) ) )^2.0 / ( #6^2.0 ) )
	}%
}

\pgfmathdeclarefunction{theoreticalLimit}{2}
{
	\pgfmathparse
	{
		#2*(#1-1.0)*(#1*1.0-(#2*(#1-1.0)+1.0))/(#1*1.0)+0*x
	}
}

\newboolean{buildtikzpics}
\setboolean{buildtikzpics}{false}

\usepackage{ulem}
\usepackage{cleveref}
\Crefname{appendix}{Appendix}{Appendices}
\Crefname{equation}{Equation}{Equations}
\Crefname{figure}{Figure}{Figures}
\Crefname{section}{Section}{Sections}
\Crefname{tabular}{Tabular}{Tabulars}
\crefname{appendix}{App.}{Apps.}
\crefname{equation}{Eq.}{Eqs.}
\crefname{figure}{Fig.}{Figs.}
\crefname{section}{Sec.}{Secs.}
\crefname{tabular}{Tab.}{Tabs.}

\newcommand{\symmps}{\textsc{SymMPS}}

\tikzset{>=stealth}

\pgfplotsset{
	compat=1.12,
	/pgf/declare function={
	    cos2(\x) = cos(deg(\x*pi));
	    g_plus(\x,\t,\s) = (\s-sqrt(\s^2+\t^2*cos2(\x)^2))/(\t*cos2(\x));
	    g_minus(\x,\t,\s) = (\s+sqrt(\s^2+\t^2*cos2(\x)^2))/(\t*cos2(\x));
	    gp(\x,\t) = ((\x - sqrt(\x^2 + \t^2))/\t); 
	    gm(\x,\t) = ((\x + sqrt(\x^2 + \t^2))/\t);
	} 
}

\pgfplotsset{%
	colormap={parula}{%
		rgb=(0.2081,0.1663,0.5292)rgb=(0.2116,0.1898,0.5777)rgb=(0.2123,0.2138,0.627)
		rgb=(0.2081,0.2386,0.6771)rgb=(0.1959,0.2645,0.7279)rgb=(0.1707,0.2919,0.7792)
		rgb=(0.1253,0.3242,0.8303)rgb=(0.0591,0.3598,0.8683)rgb=(0.0117,0.3875,0.882)
		rgb=(0.006,0.4086,0.8828) rgb=(0.0165,0.4266,0.8786)rgb=(0.0329,0.443,0.872)
		rgb=(0.0498,0.4586,0.8641)rgb=(0.0629,0.4737,0.8554)rgb=(0.0723,0.4887,0.8467)
		rgb=(0.0779,0.504,0.8384) rgb=(0.0793,0.52,0.8312)  rgb=(0.0749,0.5375,0.8263)
		rgb=(0.0641,0.557,0.824)  rgb=(0.0488,0.5772,0.8228)rgb=(0.0343,0.5966,0.8199)
		rgb=(0.0265,0.6137,0.8135)rgb=(0.0239,0.6287,0.8038)rgb=(0.0231,0.6418,0.7913)
		rgb=(0.0228,0.6535,0.7768)rgb=(0.0267,0.6642,0.7607)rgb=(0.0384,0.6743,0.7436)
		rgb=(0.059,0.6838,0.7254) rgb=(0.0843,0.6928,0.7062)rgb=(0.1133,0.7015,0.6859)
		rgb=(0.1453,0.7098,0.6646)rgb=(0.1801,0.7177,0.6424)rgb=(0.2178,0.725,0.6193)
		rgb=(0.2586,0.7317,0.5954)rgb=(0.3022,0.7376,0.5712)rgb=(0.3482,0.7424,0.5473)
		rgb=(0.3953,0.7459,0.5244)rgb=(0.442,0.7481,0.5033) rgb=(0.4871,0.7491,0.484)
		rgb=(0.53,0.7491,0.4661)  rgb=(0.5709,0.7485,0.4494)rgb=(0.6099,0.7473,0.4337)
		rgb=(0.6473,0.7456,0.4188)rgb=(0.6834,0.7435,0.4044)rgb=(0.7184,0.7411,0.3905)
		rgb=(0.7525,0.7384,0.3768)rgb=(0.7858,0.7356,0.3633)rgb=(0.8185,0.7327,0.3498)
		rgb=(0.8507,0.7299,0.336) rgb=(0.8824,0.7274,0.3217)rgb=(0.9139,0.7258,0.3063)
		rgb=(0.945,0.7261,0.2886) rgb=(0.9739,0.7314,0.2666)rgb=(0.9938,0.7455,0.2403)
		rgb=(0.999,0.7653,0.2164) rgb=(0.9955,0.7861,0.1967)rgb=(0.988,0.8066,0.1794)
		rgb=(0.9789,0.8271,0.1633)rgb=(0.9697,0.8481,0.1475)rgb=(0.9626,0.8705,0.1309)
		rgb=(0.9589,0.8949,0.1132)rgb=(0.9598,0.9218,0.0948)rgb=(0.9661,0.9514,0.0755)
		rgb=(0.9763,0.9831,0.0538)
	}
}

\newcommand{\splittedTableHeader}[2]%
{%
	\tikzset{external/export next=false}%
	\begin{tikzpicture}%
		\node[anchor=south west, inner sep = 0, outer sep = 0] (n) at (0,0) {\tiny #1};%
		\node[anchor=north east, inner sep = 0, outer sep = 0] (d) at (0,0) {\tiny #2};%
		\node[fit = (n) (d), inner sep = 0, outer sep = 0] (frame) {};%
		\draw[-] (frame.north west) -- (frame.south east);%
	\end{tikzpicture}%
}%

\newcommand{\centeredSplittedTableHeader}[2]%
{%
	\noindent\parbox[c]{\widthof{\splittedTableHeader{#1}{#2}}}%
	{\splittedTableHeader{#1}{#2}}%
}
\newacronym[shortplural={MPS}]{MPS}{MPS}{matrix\hyp product state}
\newacronym{MPO}{MPO}{matrix-product operator}
\newacronym{SVD}{SVD}{singular-value decomposition}
\newacronym{QCS}{QCS}{quantum-computer simulator}
\newacronym{FSM}{FSM}{finite-state machine}
\newacronym{ACA}{ACA}{adaptive cross approximation}
\newacronym{1D}{1D}{one\hyp dimensional}
\newacronym{QC}{QC}{quantum computer}
\newacronym{CDW}{CDW}{charge\hyp density wave}
\newacronym{SDW}{SDW}{spin\hyp density wave}
\newacronym{ARPES}{ARPES}{angle-resolved photoemission spectroscopy}
\newacronym{OBC}{OBC}{open-boundary conditions}
\newacronym{PBC}{PBC}{periodic-boundary conditions}
\newacronym{TEBD}{TEBD}{time-evolution block-decimation}
\newacronym{TDVP}{TDVP}{time\hyp dependent variational principle}
\newacronym{iff}{iff}{if and only if}
\newacronym{DFT}{DFT}{density\hyp functional theory}
\newacronym{DMFT}{DMFT}{dynamical mean\hyp field theory}
\newacronym{DMRG}{DMRG}{density\hyp matrix renormalization group}
\newacronym{1DMRG}{1DMRG}{single-site density\hyp matrix renormalization group}
\newacronym{2DMRG}{2DMRG}{two-site density\hyp matrix renormalization group}
\newacronym{DMRG3S}{DMRG3S}{strictly single-site density\hyp matrix renormalization group}
\newacronym{iDMRG}{iDMRG}{inifinite\hyp size density\hyp matrix renormalization group}
\newacronym{tDMRG}{tDMRG}{time\hyp dependent density\hyp matrix renormalization group}
\newacronym{QMC}{QMC}{quantum Monte Carlo}
\newacronym{AIM}{AIM}{Anderson impurity model}
\newacronym{SIAM}{SIAM}{single impurity Anderson model}
\newacronym{LDA}{LDA}{local\hyp density approximation}
\newacronym{LBNL}{LBNL}{Lawrence Berkeley National Laboratory}
\newacronym{VQE}{VQE}{variational\hyp quantum eigensolver}
\newacronym{ED}{ED}{exact diagonalization}
\newacronym{QPT}{QPT}{quantum phase transition}
\newacronym{QCP}{QCP}{quantum critical point}
\newacronym{ETH}{ETH}{eigenstate thermalization hypothesis}
\newacronym{EHM}{EHM}{extended Hubbard model}
\newacronym{AKLT}{AKLT}{Affleck\hyp Lieb\hyp Kennedy\hyp Tasaki}
\newglossaryentry{QR}{name={QR},description={QR decomposition}}
\newacronym{TNS}{TNS}{tensor\hyp network state}
\newacronym{SM}{SM}{supplemental material}
\newacronym{NOO}{NOO}{natural orbital occupation}
\newacronym{NO}{NO}{natural orbital}
\newacronym{LRO}{LRO}{long\hyp range order}
\newacronym{qLRO}{qLRO}{quasi\hyp long\hyp range order}
\newacronym{SC}{SC}{Superconductivity}
\newacronym{VBGS}{VBGS}{valence bond ground-state}
\newacronym{PEPS}{PEPS}{projected entangled pair\hyp states}
\newacronym{ALS}{ALS}{alternating least squares}
\newacronym{BdG}{BdG}{Bogoljubov de-Gennes}
\newacronym{TFIM}{TFI}{transverse field Ising model}
\newacronym{PP}{PP}{projected purification}
\newacronym{BEC}{BEC}{Bose\hyp Einstein condensate}
\newacronym{JWT}{JWT}{Jordan\hyp Wigner transformation}
\newacronym{NISQ}{NISQ}{noisy intermediate scale quantum}
\newacronym{NN}{NN}{nearest\hyp neighbor}
\newacronym{NNN}{NNN}{next\hyp nearest\hyp neighbor}
\newacronym{SPDM}{SPDM}{single\hyp particle density matrix} 
\newacronym{HCB}{HCB}{hardcore bosons}
\newacronym{SF}{SF}{spinless fermions}%
\begin{document}%
\title{Symmetry\hyp protected Bose\hyp Einstein condensation of interacting hardcore bosons}%
\author{R.~H.~Wilke}%
\affiliation{Department of Physics, Arnold Sommerfeld Center for Theoretical Physics (ASC), Ludwig-Maximilians-Universit\"{a}t M\"{u}nchen, 80333 M\"{u}nchen, Germany}%
\author{T.~K\"ohler}%
\affiliation{Department of Physics and Astronomy, Uppsala University, Box 516, S-751 20 Uppsala, Sweden}%
\author{F.~A.~Palm}%
\affiliation{Department of Physics, Arnold Sommerfeld Center for Theoretical Physics (ASC), Ludwig-Maximilians-Universit\"{a}t M\"{u}nchen, 80333 M\"{u}nchen, Germany}%
\author{S.~Paeckel}%
\affiliation{Department of Physics, Arnold Sommerfeld Center for Theoretical Physics (ASC), Ludwig-Maximilians-Universit\"{a}t M\"{u}nchen, 80333 M\"{u}nchen, Germany}%
\date{\today}%
\begin{abstract}%
We introduce a mechanism stabilizing a one\hyp dimensional quantum many\hyp body phase, characterized by a certain wave vector $k_0$, from a $k_0$\hyp modulated coupling to a center site, via the protection of an emergent $\mathbb Z_2$ symmetry. %
We illustrate this mechanism by constructing the solution of the full quantum many\hyp body problem of hardcore bosons on a wheel geometry, which are known to form a \acrlong{BEC}. %
The robustness of the condensate is shown numerically by adding \acrlong{NN} interactions to the wheel Hamiltonian. %
We identify the energy scale that controls the protection of the emergent $\mathbb Z_2$ symmetry. %
We discuss further applications such as geometrically inducing finite\hyp momentum condensates. %
Since our solution strategy is based on a generic mapping from a wheel geometry to a projected ladder, our analysis can be applied to various related problems with extensively scaling coordination numbers.
\end{abstract}%
\glsresetall%
\maketitle%
Cold atom experiments have become a versatile platform to realize various exotic quantum phases of matter~\cite{supersolid_penrose,Anderson198,Jaksch1998,Bloch2008,dalfovo99,review_natphys_giamarchi,deEscobar:2009p2001,Kraft:2009p2011,Bloch2008}. %
Available experimental setups nowadays allow for the control of both geometry and interactions of simulated model systems. %
It is thus crucial to theoretically identify physical mechanisms that improve the stability and scaling properties of exotic quantum phases, which then might be realized and tested in experiments.
In that context, remarkable progress in understanding the stability of \gls{BEC} has been made by analyzing spectral properties of a wheel of \gls{HCB}~\cite{Dongen1991,Vidal2011,Tennie2017,Mate2021} as depicted in~\cref{fig:geometry:wheel}. %
This model features an energy scale $\sim \sqrt L$ that is generated by the extensively scaling coordination number of a center site. %
While large coordination numbers appear in several theoretical approaches~\cite{wilson_rg,georges_RMP,Freericks2006,PhysRevB.88.075135,Aoki2014,Seth2016}, the exact solution as well as the stability against perturbations remained an open question. %
Besides others, the problem of finding exact expressions for ground states of long\hyp range coupled \gls{HCB} Hamiltonians is a major obstacle. %
Here, arbitrarily long\hyp ranged interactions appear when expressing the \gls{HCB} degrees of freedom in terms of \gls{SF} via a~\gls{JWT}. %
\begin{figure}%
	\centering%
	\ifthenelse{\boolean{buildtikzpics}}%
	{%
		\tikzsetnextfilename{dispersion_relation}%
		\begin{tikzpicture}%
			\clip (-0.18,0) rectangle + (8.6,5.525);%
			\begin{axis}%
			[%
				width			=	0.55\textwidth,%
				height			=	0.275\textheight,%
				axis y line		=	middle,%
				axis x line 	=	middle,%
				xtick			=	{-1,-0.5, 0.5, 1},%
				ytick			=	{-2,2,6,10},%
				xmin			=	-1,%
				xmax			=	1,%
				ymin			=	-5.25,%
				ymax			=	14.05,%
				xlabel			=	{\normalsize $k\left[\frac{\pi}{a}\right]$},%
				xlabel style	=	{xshift=0.25cm, yshift=0cm},%
				ylabel			=	{\normalsize $E[t]$},%
				ylabel style	=	{xshift=0cm, yshift=0.25cm},%
			]%
				\addplot [orange, densely dotted, thick, samples = 2000, opacity=0.5] {cos2(x)+sqrt(cos2(x)^2 + 4^2)};%
				\addplot [colorC, densely dotted, thick, samples = 2000, opacity=0.5] {cos2(x)-sqrt(cos2(x)^2 + 4^2)};%
				\addplot [red!75!black, thick, samples = 2000] {2*cos2(x)};%
				\node [circle, draw=red, fill=white, minimum height=3.5pt, inner sep=0pt, thick] at (axis cs: 0,2) {};%
				\pgfmathparse{cos2(0)+sqrt(cos2(0)+4^2)}\edef\ekup{\pgfmathresult}%
				\node [cross out, draw=red, thick, inner sep=2pt] at (axis cs: 0, \ekup) {};%
				\node [text=colorG!70!black, font=\large] at (axis cs: -.1, \ekup) {$\varepsilon_+$};%
				\pgfmathparse{cos2(0)-sqrt(cos2(0)+4^2)}\edef\ekdown{\pgfmathresult}%
				\node [cross out, draw=red, thick, inner sep=2pt] at (axis cs: 0, \ekdown) {};%
				\node [text=colorG!70!black, font=\large] at (axis cs: -.1, \ekdown) {$\varepsilon_-$};%
				\draw [->, thick, draw=colorC!70!black, thick, bend right] (axis cs: 0,2) to node[midway, right] {$\propto \tilde s$} (axis cs: 0,\ekup);%
				\draw [->, thick, draw=colorC!70!black, thick, bend left] (axis cs: 0,2) to node[midway, right] {$\propto \tilde s$} (axis cs: 0,\ekdown);%
				\foreach \kval in {-.66,-.33}%
				{%
					\pgfmathparse{cos2(\kval)+sqrt(cos2(\kval)+4^2)}\edef\ekladderup{\pgfmathresult}%
					\pgfmathparse{cos2(\kval)-sqrt(cos2(\kval)+4^2)}\edef\ekladderdown{\pgfmathresult}%
					\pgfmathparse{2*cos2(\kval)}\edef\ekwheel{\pgfmathresult}%
					\edef\tmp%
					{%
						\noexpand\draw [->, thick, draw=orange!70, thick, opacity=0.5] (axis cs: \kval,\ekladderup) to (axis cs: \kval,\ekwheel);%
						\noexpand\draw [->, thick, draw=colorC!70, thick, opacity=0.5] (axis cs: \kval,\ekladderdown) to (axis cs: \kval,\ekwheel);%
					}%
					\tmp%
				}%
				\foreach \kval in {.66,.33}%
				{%
					\pgfmathparse{cos2(\kval)+sqrt(cos2(\kval)+4^2)}\edef\ekladderup{\pgfmathresult}%
					\pgfmathparse{cos2(\kval)-sqrt(cos2(\kval)+4^2)}\edef\ekladderdown{\pgfmathresult}%
					\pgfmathparse{2*cos2(\kval)}\edef\ekwheel{\pgfmathresult}%
					\edef\tmp%
					{%
						\noexpand\draw [->, thick, draw=orange!70, thick, opacity=0.5] (axis cs: \kval,\ekladderup) to (axis cs: \kval,\ekwheel);%
						\noexpand\draw [->, thick, draw=colorC!70, thick, opacity=0.5] (axis cs: \kval,\ekladderdown) to (axis cs: \kval,\ekwheel);%
					}%
					\tmp%
				}%
				\node [text=colorG!70!black, font=\large] at (axis cs: .75, -2.) {$\varepsilon_k$};%
				\coordinate (leftCircle) at (axis description cs:0.175,0.825);%
				\coordinate (rightCircle) at (axis description cs:0.825,0.825);%
			\end{axis}%
			\def\nmax{13}%
			\def\radius{1.25}%
			\node[circle, draw, inner sep = 0, minimum width = 0.5em, line width = 0.02em] at (leftCircle) (center) {};%
			\foreach \n in {1,...,\nmax}%
			{%
				\node[circle, draw, inner sep = 0, minimum width = 0.5em, line width = 0.02em] at ($({\radius*cos((\n-1)/\nmax*360)},{\radius*sin((\n-1)/\nmax*360)})+(center)$) (ring\n) {};%
				\draw[densely dashed, draw=red!70!black, line width = 0.02em] (center) to (ring\n);%
				\node at ($({0.7*\radius*cos((\n-1+0.3)/\nmax*360)},{0.7*\radius*sin((\n-1+0.3)/\nmax*360)})+(center)$) {$s$};%
			}%
			\pgfmathparse{\nmax-1}%
			\def\nmaxmm{\pgfmathresult}%
			\foreach \n in {1,...,\nmaxmm}%
			{%
				\pgfmathparse{int(\n+1)}%
				\edef\npp{\pgfmathresult}%
				\draw[densely dotted, line width = 0.02em] (ring\n) to (ring\npp);%
				\node at ($({1.1*\radius*cos((\n-1+0.5)/\nmax*360)},{1.1*\radius*sin((\n-1+0.5)/\nmax*360)})+(center)$) {$t$};%
			}%
			\draw[densely dotted, line width = 0.02em] (ring\nmax) to (ring1);%
			\node at ($({1.1*\radius*cos((-0.5)/\nmax*360)},{1.1*\radius*sin((-0.5)/\nmax*360)})+(center)$) {$t$};%
			\node[draw, anchor = south east] at ($({1.6*\radius*cos(135)},{1.6*\radius*sin(135)})+(center)+(0,0.2)$) {\subfloat[\label{fig:geometry:wheel}]{}};%
			\def\InnerRadius{0.7}
			\def\OuterRadius{1.25}
			\node[] at (rightCircle) (center) {};%
			\foreach \n in {1,...,\nmax}%
			{%
				\node[circle, draw, inner sep = 0, minimum width = 0.5em, line width = 0.02em] at ($({\InnerRadius*cos((\n-1)/\nmax*360)},{\InnerRadius*sin((\n-1)/\nmax*360)})+(center)$) (InnerRing\n) {};%
				\node[circle, draw, inner sep = 0, minimum width = 0.5em, line width = 0.02em] at ($({\OuterRadius*cos((\n-1)/\nmax*360)},{\OuterRadius*sin((\n-1)/\nmax*360)})+(center)$) (OuterRing\n) {};%
				\draw[dashed, draw=red!70!black, line width = 0.02em] (InnerRing\n) to (OuterRing\n);%
				\node at ($({((\OuterRadius+\InnerRadius)/2.0)*cos((\n-1+0.3)/\nmax*360)},{((\OuterRadius+\InnerRadius)/2.0)*sin((\n-1+0.3)/\nmax*360)})+(center)$) {$s$};%
			}%
			\pgfmathparse{\nmax-1}%
			\def\nmaxmm{\pgfmathresult}%
			\foreach \n in {1,...,\nmaxmm}%
			{%
				\pgfmathparse{int(\n+1)}%
				\edef\npp{\pgfmathresult}%
				\draw[densely dotted, line width = 0.02em] (OuterRing\n) to (OuterRing\npp);%
				\node at ($({1.1*\OuterRadius*cos((\n-1+0.5)/\nmax*360)},{1.1*\OuterRadius*sin((\n-1+0.5)/\nmax*360)})+(center)$) {$t$};%
			}%
			\draw[densely dotted, line width = 0.02em] (OuterRing\nmax) to (OuterRing1);%
			\node at ($({1.1*\OuterRadius*cos((-0.5)/\nmax*360)},{1.1*\OuterRadius*sin((-0.5)/\nmax*360)})+(center)$) {$t$};%
			\node[draw, anchor = south east] at ($({1.6*\radius*cos(135)},{1.6*\radius*sin(135)})+(center)+(0,0.2)$) {\subfloat[\label{fig:geometry:ladder}]{}};%
		\end{tikzpicture}%
	}%
	{%
		{\textcolor{white}{\subfloat[\label{fig:geometry:wheel}]{}\subfloat[\label{fig:geometry:ladder}]{}}}
		\includegraphics{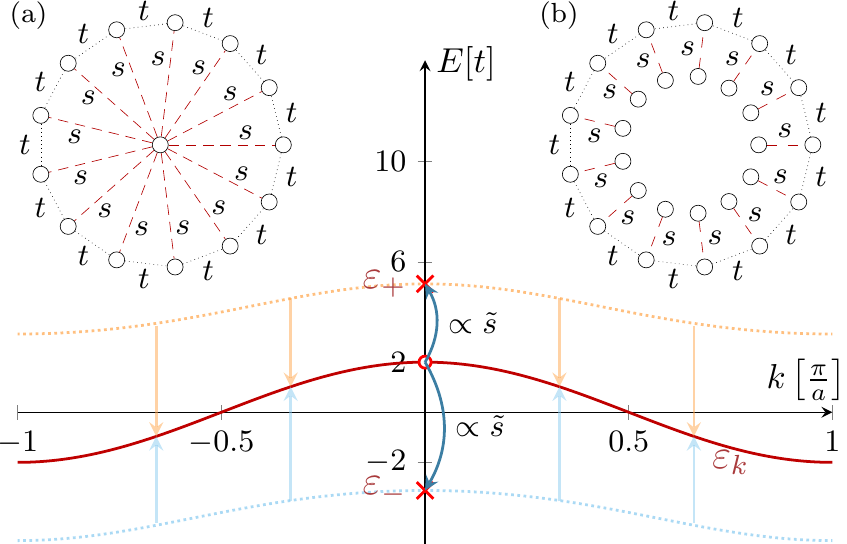}%
	}%
	\caption%
	{%
		\label{fig:wheel-to-ladder:sp-dispersion}%
		The main plot illustrates the single\hyp particle dispersion relation (middle, red curve) of the wheel geometry \protect\subref{fig:geometry:wheel}, emerging from projecting down the dispersion from the ladder geometry \protect\subref{fig:geometry:ladder} (upper orange and lower blue curve). %
		Note the appearance of two single\hyp particle states at $k_0=0$ (red crosses). %
		This is because the Hilbert space of the wheel is obtained by projecting out all modes on the inner ring, except for the zero momentum states $\ket{N_{\odot, k=0}}_\odot$.
		Momentum conservation then couples this central mode to the particular mode on the outer ring respecting the $k_0$\hyp modulated ring\hyp to\hyp center hopping, which generates an extensively scaling level splitting (red circle and crosses). %
	}%
\end{figure}
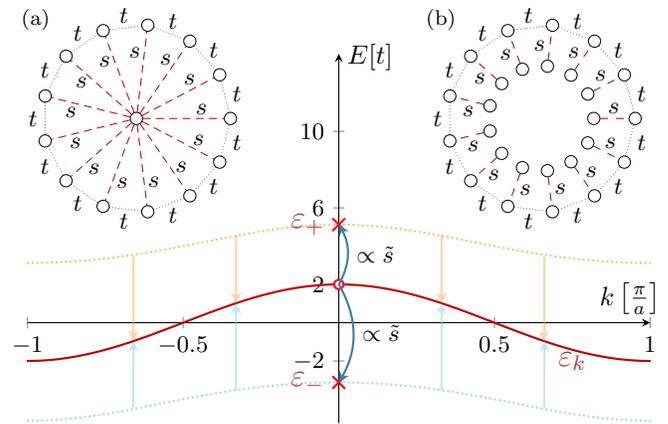%
In this letter, we present a mapping that allows us to construct the full solution of a family of quantum many\hyp body problems with arbitrary $k_0$\hyp modulated ring\hyp to\hyp center hoppings $s_j = s\mathrm{e}^{\mathrm{i}k_0 j}$, and to analyze the formation of a \gls{BEC} phase with momentum $k_0$. %
In the context of central spin models~\cite{Gaudin1976,Prokofextquotesingleev2000,Delft2001,Taylor2003,Dukelsky2004,Villazon2020} a solution strategy to a similar problem is based on the Bethe ansatz and has been applied to describe for instance Rydberg impurities in ultracold atomic quantum gases~\cite{Ashida2019}. %
In contrast, we derive the solution by introducing a mapping to a ladder system of \gls{SF}, yielding closed analytical expressions. %
We emphasize that this mapping can be applied in various other setups to analytically tackle problems with an extensively scaling coordination number.
In the context of hardcore bosons, our approach reveals that the stabilizing mechanism for the \gls{BEC} is the extensively scaling coordination number of the center site, introducing a robust discrete $\mathbb Z_2$ symmetry that protects the ordered quantum many\hyp body phase against local perturbations on the outer ring. %
Furthermore, we trace back the protection to odd\hyp parity $k=k_0$ single\hyp particle states that are gapped out $\sim  s\sqrt L \equiv \tilde s$. %
This property allows us to show that in the thermodynamic limit the system immediately transitions into a \gls{BEC}, as long as there is a finite ring\hyp to\hyp center hopping rate $s>0$, which remarkably also holds when adding local interactions to the outer ring. %
We demonstrate, beyond previous work, the robust protection of the~\gls{BEC} numerically, using~\gls{DMRG}~\cite{Schollwock:2005p2117,Schollwoeck201196} simulations to calculate the $k_0$\hyp condensate fraction when adding \gls{NN} interactions, for a wide parameter range and various particle densities. %
As a consequence, the $\mathbb Z_2$ symmetry in principle allows to experimentally tune the transition temperature of a gas of interacting \gls{HCB} by modifying the wheel's coordination number. %
Here, we show that the central quantity is the ratio $\frac{V}{\tilde s}$ between the interaction strength $V$ and the renormalized ring\hyp to\hyp center hopping, which we demonstrate by further numerical results. %
Finally, our analysis implies that the emergent $\mathbb Z_2$ symmetry is generically induced by the model's geometry. %
Therefore, general $k_0$\hyp modulated hoppings give rise to corresponding protected $k_0$\hyp modes and the respective single\hyp particle states are gapped out $\sim s\sqrt L$. %
This paves the way to a generic mechanism that can be exploited in various contexts, for instance, to stabilize exotic quantum many\hyp body phases such as $k_0\neq 0$ \gls{BEC}~\cite{Lim2008,Hick2010,DiLiberto2011,Lin2011}. %
\paragraph{\label{sec:wheel-to-ladder-mapping}Analytical solution via wheel-to-ladder mapping.---}%

We consider \gls{HCB} on a $L$\hyp sited ring with an additional center site~\cite{Vidal2011,Mate2021} (see \cref{fig:geometry:wheel}).
The model exhibits $k_0$\hyp modulated ring\hyp to\hyp center hopping $s_j = s\mathrm{e}^{\mathrm{i}k_0 j}$ while the homogeneous hopping on the ring is tuned by a parameter $t$.
The corresponding Hamiltonian reads %
\begin{align}%
	\hat H%
	\equiv%
	-t \sum_{j=0}^{L-1} \left( \hat h^\dagger_j \hat h^\nodagger_{j+1} + \mathrm{h.c.} \right) -%
	\sum_{j=0}^{L-1} \left( s_j\hat h^\dagger_j \hat h^\nodagger_\odot + \mathrm{h.c.} \right) \; ,%
	\label{eq:hubb-wheel}%
\end{align}%
where $\hat h^{(\dagger)}_j$ is the \gls{HCB} ladder operator on the $j$-th site of the ring and $\hat h^{(\dagger)}_\odot$ on the center site, spanning the overall Hilbert space $\mathcal H^{\otimes L+1}_2$. %
In the limit $\frac{s}{t} \rightarrow 0$ (ring geometry) the model exhibits a quasi \gls{BEC}, i.e., the ground state is a bosonic condensate, whose occupation scales as $\sqrt N$~\cite{Lieb1963,Lieb1963a}, where $N$ denotes the number of \gls{HCB}.
The opposite limit, $\frac{s}{t}\rightarrow \infty$ (star geometry), has been shown recently to feature a real BEC where the occupation in the ground state scales as $L \rho \left(1-\rho + 1/L\right)$ with $\rho=\nicefrac{N}{L}$~\cite{Tennie2017}. %
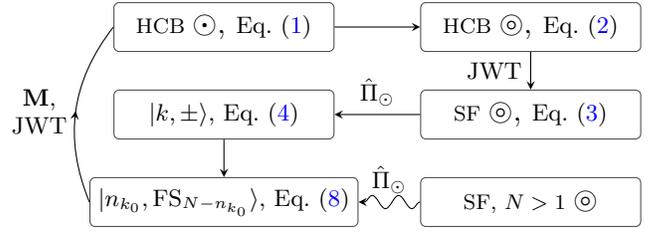
\begin{figure}%
	\centering%
	\ifthenelse{\boolean{buildtikzpics}}%
	{%
		\tikzsetnextfilename{solution_strategy}%
		\begin{tikzpicture}%
			[%
			block/.style = {draw=black!70, rounded corners=2pt, font=\footnotesize, minimum width=9em, minimum height=2em},%
			node distance = 0.5 and 0.85,
			]%
			\node [block] (hcb_wheel) {HCB \large $\odot$, \small\cref{eq:hubb-wheel}};%
			\node [block, right=3.5em of hcb_wheel] (hcb_ladder) {HCB \large $\circledcirc$, \small\cref{eq:wheel-ladder}};%
			\node [block, below=of hcb_ladder] (fsp_ladder) {SF \large $\circledcirc$, \small\cref{eq:hubb-ladder}};%
			\node [block, below=of fsp_ladder] (fmp_ladder) {SF, $N>1$ \large $\circledcirc$};%
			\node [block] at (hcb_wheel|-fmp_ladder) (fmp_states_wheel) {\small $\ket{n_{k_0},\mathrm{FS}_{N-n_{k_0}}}$, \small\cref{eq:hubb-wheel:bosonic:mb-eigenstates}};%
			\node [block] at (hcb_wheel|-fsp_ladder) (fsp_states_wheel) {\small $\ket{k, \pm}$, 	\cref{eq:hubb-ladder:fermionic:sp-eigenstates}};%
			\draw [->] (hcb_wheel) to (hcb_ladder);%
			\draw [->] (hcb_ladder) to node[midway, left] {\gls{JWT}} (fsp_ladder);%
			\draw [->] (fsp_ladder) to node[midway, above] {$\hat \Pi_\odot$} (fsp_states_wheel);%
			\draw [orientedsnake] (fmp_ladder) to node[midway, above] {$\hat \Pi_\odot$} (fmp_states_wheel);%
			\draw [->] (fsp_states_wheel) to (fmp_states_wheel);%
			\draw [->-, bend left] (fmp_states_wheel.west) to node[midway, left, align=center] {$\mathbf M$, \\ \gls{JWT}} (hcb_wheel.west);%
		\end{tikzpicture}%
	}%
	{%
		\includegraphics{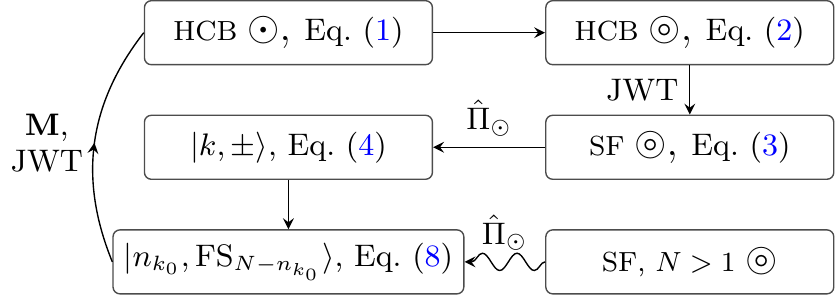}%
	}%
	\caption%
	{%
		\label{fig:solution-strategy}%
		Solution strategy for~\cref{eq:hubb-wheel}. %
		The \gls{HCB} wheel \begin{large}$\odot$\end{large} is transformed to a ladder \begin{large}$\circledcirc$\end{large}, which is then mapped to a ladder of spinless \gls{SF} via a \gls{JWT}. %
		From the ladder of SF, the single\hyp particle spectrum $\ket{k,\pm}$ and therefrom projected Slater determinants $\ket{n_{k_0},\mathrm{FS}_{N-n_{k_0}}}$ are constructed, utilizing the projector to the $N_\odot \leq 1$ subspace, $\hat \Pi_\odot$.
		The constructed many\hyp particle Slater determinants finally allow for the analytic solution of the full \gls{HCB} wheel diagonalizing a $4\times 4$ matrix $\mathbf M$. %
		Note that no closed solution of the SF ladder Hamiltonian is required (only its projected counterpart). %
	}%
\end{figure}%
In order to construct a full analytical solution, we introduce a mapping from the wheel~\cref{eq:hubb-wheel} to a ladder geometry of \gls{HCB} with periodic boundary conditions (see~\cref{fig:geometry:ladder}). %
The overall solution strategy is sketched schematically in~\cref{fig:solution-strategy}. %
The crucial step is to identify the central Hilbert space of the \gls{HCB} wheel with the subspace of the single\hyp particle momentum states $\ket{N_{\odot,k=0}}_\odot$ on the inner ring of the ladder (enforcing occupations $N_{\odot,k=0} \equiv N_\odot \leq 1$).
The projector $\hat \Pi_\odot$ to this subspace allows us to construct a solution on the expanded Hilbert space of the ladder geometry and eventually project down.
Thereby, the long\hyp range coupled wheel Hamiltonian can be mapped to an only \gls{NNN} coupled ladder Hamiltonian $\hat H_{\mathrm{lad}}$:
\begin{align}%
	\hat H = \hat \Pi_\odot \hat H_{\mathrm{lad}} \hat \Pi_\odot\; .%
	\label{eq:wheel-ladder}%
\end{align}%
While the full details of the mapping can be found in~\cite{suppMat}, the most important observation is that a \gls{JWT} of $\hat H_{\mathrm{lad}}$ introduces only local parity operators $\mathrm{e}^{\mathrm{i} \pi \hat n^\nodagger_{\odot, j}}$: %
\begin{align}%
	\hat \Pi_\odot \hat H_{\mathrm{lad}} \hat \Pi_\odot%
	&=%
	t \sum_j \hat \Pi_\odot \left( \hat c^\dagger_j \mathrm{e}^{\mathrm{i} \pi \hat n^\nodagger_{\odot, j}} \hat c^\nodagger_{j+1} + \mathrm{h.c.} \right) \hat \Pi_\odot \notag \\%
	&\phantom{=}%
	-%
	\tilde s \sum_j \hat \Pi_\odot \left( \mathrm{e}^{\mathrm{i}k_0 j} \hat c^\dagger_j \hat c^\nodagger_{\odot,j} + \mathrm{h.c.} \right) \hat\Pi_\odot \; ,%
	\label{eq:hubb-ladder}%
\end{align}%
wherein $\hat c^{(\dagger)}_j$ $(\hat c^{(\dagger)}_{\odot,j})$  denotes the fermionic ladder operator on the $j$-th site of the outer (inner) ring and the single\hyp site number operator on the inner ring is given by $\hat n^\nodagger_{\odot, j} = \hat c^\dagger_{\odot, j} \hat c^\nodagger_{\odot, j}$. %
We emphasize the appearance of a rescaled ring\hyp to\hyp center hopping amplitude $\tilde s = s \sqrt L$, that allows to connect to the known solutions when taking the thermodynamic limit $L\rightarrow\infty$. %
In fact, in the thermodynamic limit, the wheel immediately collapses to the star geometry whenever there is a fixed, finite ratio $\frac{s}{t}$, and the ground state is a true \gls{BEC}. %
However, the question remains what happens for fixed ratios $\frac{\tilde s}{t}$. %
This matters for finite system sizes, as is the case for mesoscopic systems, in ultracold atomic gas experiments or Rydberg atoms~\cite{Eiles2021}. %
In particular, we are interested in the impact of the extensive energy scale set by $\tilde s$ on the formation and stability of the \gls{BEC}, which requires a more in\hyp depth analysis of the ground state of~\cref{eq:hubb-ladder}. %
Note that for now and in the following, we refer to the scaling of the ring\hyp to\hyp center hopping $\tilde s = s\sqrt L$ as extensive in the system size. %
It is instructive to first solve~\cref{eq:hubb-ladder} for the single\hyp particle eigenstates 	$\ket{k, \pm}$, fulfilling $\braket{\mathrm{e}^{\mathrm{i} \pi \hat n^\nodagger_{\odot, j}}} \equiv 1$:
\begin{align}%
	\ket{k,\pm}%
	&=
	\begin{cases}%
	c^\dagger_k \ket{\varnothing} &\text{if $k\neq k_0$,} \\%
	\psi_\pm \left( \hat c^\dagger_k + \Delta_{\pm} \hat c^\dagger_{\odot, k=0}\right) \ket{\varnothing} &\text{if $k= k_0$.}%
	\end{cases}%
	\label{eq:hubb-ladder:fermionic:sp-eigenstates}%
\end{align}%
Here, $\hat c^\dagger_{(\odot), k} = \frac{1}{\sqrt L} \sum\limits_{j=0}^{L-1} \mathrm{e}^{-\mathrm{i} kj} \hat c^\dagger_{(\odot),j}$ with $\psi_{(k),\pm}$ being a normalization constant and $\ket{\varnothing}$ denotes the vacuum state.
As shown in~\cref{fig:wheel-to-ladder:sp-dispersion}, the corresponding single\hyp particle spectrum is identical to that of a tight\hyp binding chain (i.e., $\varepsilon_k = 2t \cos k$) except for the $k = k_0$ states whose single\hyp particle energies are characterized by the splitting $\Delta_\pm = \frac{\varepsilon_0}{2\tilde s} \pm \frac{\sqrt{\varepsilon^2_0 + 4\tilde s^2}}{\lvert 2\tilde s \rvert}$:
\begin{align}%
	\varepsilon_\pm = \frac{1}{2}\left(\varepsilon_0 \pm \operatorname{sgn}(\tilde s)\sqrt{\varepsilon^2_0 + 4 \tilde s^2} \right) = \tilde s \Delta_\pm \; .%
	\label{eq:hubb-ladder:fermionic:sp-energies:k=0}%
\end{align}%
These $k=k_0$ single\hyp particle eigenstates~\cref{eq:hubb-ladder:fermionic:sp-eigenstates} separate $\propto \lvert \tilde s\rvert \propto \sqrt L$ from the remaining spectrum giving rise to a single\hyp particle gap. %
Referring to~\cref{eq:hubb-ladder}, in the limit $\frac{\tilde s}{t} \rightarrow \infty$, the hopping on the outer ring can be neglected, and the same holds for the impact of the \gls{JWT} on the overall eigenstate. %
Consequently, the single\hyp particle gap can be expected to control the many\hyp body spectrum, in this limit.
Additionally, from $\Delta_\pm \stackrel{\nicefrac{\tilde s}{t} \rightarrow \infty}{\longrightarrow} \pm 1$ we find that the corresponding wavefunction is characterized by a maximally mixing of the distinguished mode $\ket{k_0}$ on the outer ring with the state $\ket{N_\odot=1}_\odot$ on the inner ring. %
This highly non\hyp local wavefunction, generated from the extensive scaling of~\cref{eq:hubb-ladder:fermionic:sp-energies:k=0}, already suggests the stability of the \gls{BEC} under local perturbations on the outer ring.
In order to further elaborate on the extensive scaling property, we now return to the full solution of~\cref{eq:hubb-wheel} with the complete derivation detailed in~\cite{suppMat}. %
Here, the key observation is that Slater determinants constructed from single\hyp particle states~\cref{eq:hubb-ladder:fermionic:sp-eigenstates} with $k\neq k_0$ are also eigenstates of $\hat H = \hat \Pi_\odot \hat H_{\mathrm{lad}} \hat \Pi_\odot$:%
\begin{align}%
	\hat \Pi_\odot \hat H_{\mathrm{lad}} \hat \Pi_\odot \ket{\mathrm{FS}_N} = \left(\sum_{l=1}^{N} \varepsilon_{k_l}\right) \hat \Pi_\odot \ket{\mathrm{FS}_N} \; ,%
\end{align}%
where $\ket{\mathrm{FS}_N} = \ket{ k_1, \ldots, k_N }$ denotes a Slater determinant labeled by a set of $N$ single\hyp particle eigenstates with $k_l \neq k_0$. %
This observation can be understood by noting that the projected parity operator in~\cref{eq:hubb-ladder} can be written in terms of the zero momentum density $N_\odot$ on the inner ring
\begin{align}
	\hat \Pi_\odot e^{\mathrm i \pi \hat n_{\odot,j}}\hat \Pi_\odot = \hat 1_{\odot,j} - \frac{2}{L} \hat c^\dagger_{\odot,k=0}\hat c^\nodagger_{\odot,k=0} \; ,
\end{align}
and thus $\hat \Pi_\odot e^{\mathrm i \pi \hat n_{\odot,j}} \hat \Pi_\odot \ket{\mathrm{FS}_N} = \ket{\mathrm{FS}_N}$.
Particle\hyp number conservation of the wheel Hamiltonian then motivates to construct an ansatz for the $N$\hyp particle eigenstates, superimposing all possible occupations of the $k_0$ mode that belong to the same overall particle number sector
\begin{align}%
	\ket{\psi_N}%
	&=%
	\alpha_0 \ket{\mathrm{FS}_N}%
	+%
	\left( \alpha_{1_+} \hat \psi^\dagger_{+} + \alpha_{1_-} \hat \psi^\dagger_{-} \right) \ket{\mathrm{FS}_{N-1}} \notag \\%
	&\phantom{= \alpha_0 \ket{\mathrm{FS}_N}+\alpha_0 \ket{\mathrm{FS}_N}}%
	+%
	\alpha_2 \hat \psi^\dagger_{+}\hat \psi^\dagger_{-} \ket{\mathrm{FS}_{N-2}} \;,%
	\label{eq:hubb-wheel:bosonic:mb-eigenstates}%
\end{align}%
with complex coefficients $\alpha_{0,1_\pm,2}$. %
These states describe a superposition of either empty ($\propto \alpha_0$) or doubly occupied ($\propto \alpha_2$) $k_0$ states and highly non\hyp local states $\propto \alpha_{1_\pm}$ in which the $k_0$ mode on the outer ring is coupled to the $\ket{N_\odot=1}_\odot$ mode on the inner ring.
\begin{figure}%
	\centering%
	\ifthenelse{\boolean{buildtikzpics}}%
	{%
		\tikzsetnextfilename{mp_eigenvalues_mt}%
		\begin{tikzpicture}%
			\begin{axis}%
			[%
				width				=	0.495\textwidth-4.79pt,%
				height				=	0.25\textheight,%
				xmode				=	log,%
				ymin				=	-105,%
				ymax				=	49,%
				xmin				=	0.01,%
				xmax				=	10,%
				xlabel				=	{$s [t]$},%
				ylabel				=	{$E_n [t]$},%
				ylabel style		=	{yshift=-1em},%
				legend columns		=	2,%
				legend entries		=	{$E_0$, $E_{1+}$, $E_{1-}$, $E_2$},%
				legend pos			=	{north west},%
				legend style		=	{draw=none},%
				unbounded coords	=	jump,%
				mark				=	none,%
			]%
				\addlegendimage{color=colorA, thick}%
				\addlegendimage{color=colorB, thick}%
				\addlegendimage{color=colorC, thick}%
				\addlegendimage{color=colorD, thick}%
				\foreach \k in {1,...,70}%
				{%
					\addplot%
					[%
						color	=	colorA
					]%
						table%
						[%
							x expr	= \thisrowno{0},%
							y expr	= \thisrowno{\k},%
						]%
							{../ed/python/mb-spectrum.L-10.N-4};%
				}%
				\foreach \k in {71,73,...,181}%
				{%
					\addplot%
					[%
						color	=	colorB,%
					]%
						table%
						[%
							x expr	= \thisrowno{0},%
							y expr	= \thisrowno{\k},%
						]%
							{../ed/python/mb-spectrum.L-10.N-4};%
				}%
				\foreach \k in {72,74,...,182}%
				{%
					\addplot%
					[%
						color	=	colorC,%
					]%
						table%
						[%
							x expr	= \thisrowno{0},%
							y expr	= \thisrowno{\k},%
						]%
							{../ed/python/mb-spectrum.L-10.N-4};%
				}%
				\foreach \k in {183,...,210}%
				{%
					\addplot%
					[%
						color	=	colorD,%
					]%
						table%
						[%
							x expr	= \thisrowno{0},%
							y expr	= \thisrowno{\k},%
						]%
							{../ed/python/mb-spectrum.L-10.N-4};%
				}%
				\draw[<->, thick, draw=black!70] (axis cs: 2,-24) to node[midway, right] {$\Delta_{1}$} (axis cs: 2,-5.5);%
				\draw[<->, thick, draw=black!70] (axis cs: 4,-36) to node[midway, right] {$\Delta_{2}$} (axis cs: 4,-5.5);%
				\coordinate (insetPos) at (axis cs: 0.019, -20);%
			\end{axis}%
			\begin{axis}%
			[%
				at					=	(insetPos),%
				anchor				=	north west,%
				width				=	0.325\textwidth,%
				height				=	0.15\textheight,%
				xmode				=	log,%
				xmin				=	0.1,%
				xmax				=	2.0,%
				mark				=	none,%
				unbounded coords	=	jump,%
				xticklabel style	=	{font=\scriptsize},%
				yticklabel style	=	{font=\scriptsize},%
			]%
				\foreach \k in {1,...,70}%
				{%
					\addplot%
					[%
						color	= colorA,%
					]%
						table%
						[%
							x expr	= \thisrowno{0},%
							y expr	= \thisrowno{\k},%
						]%
							{../ed/python/mb-spectrum.L-10.N-4};%
				}%
				\foreach \k in {72,74,...,182}%
				{%
					\addplot%
					[%
						color	= colorC,%
					]%
						table%
						[%
							x expr	= \thisrowno{0},%
							y expr	= \thisrowno{\k},%
						]%
							{../ed/python/mb-spectrum.L-10.N-4};%
				}%
				\foreach \k in {183,...,210}%
				{%
					\addplot%
					[%
						color	= colorD,%
					]%
						table%
						[%
							x expr	= \thisrowno{0},%
							y expr	= \thisrowno{\k},%
						]%
							{../ed/python/mb-spectrum.L-10.N-4};%
				}%
				\draw[-, dashed, thick, draw=black!70] (axis cs: 0.24,-30) to node[midway, left, yshift=-1.9em] {$s_{\mathrm{c},1}$} (axis cs: 0.24,10);%
				\draw[-, dashed, thick, draw=black!70] (axis cs: 0.99,-30) to node[midway, right, yshift=-1.9em] {$s_{\mathrm{c},2}$} (axis cs: 0.99,10);%
			\end{axis}%
		\end{tikzpicture}%
	}%
	{%
		\includegraphics{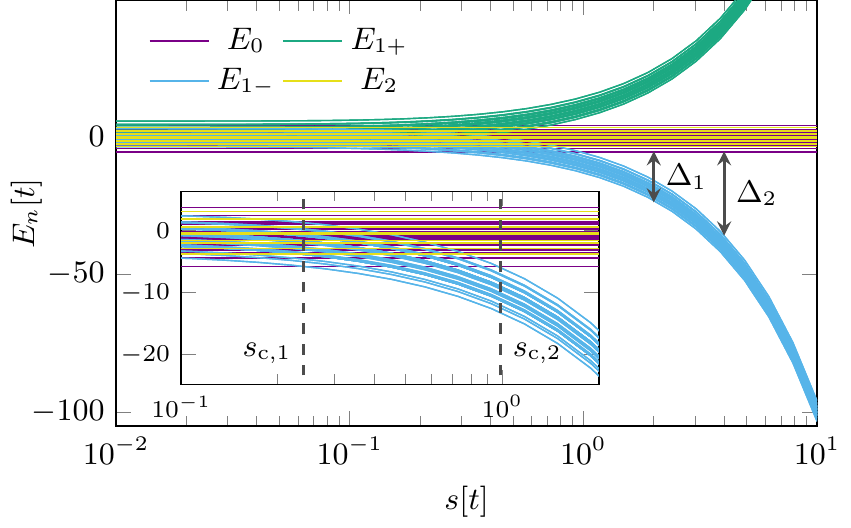}%
	}%
	\vspace{-1em}%
	\caption%
	{%
		\label{fig:mp-eigenvalues}%
		Clustering of the many\hyp particle eigenstates for a wheel composed of $10$ lattice sites in the $N=4$ particle number sector as a function of the ring\hyp to\hyp center hopping $s$. %
		Different colors correspond to the clustered energies generated from the different eigenvalues of ~\cref{eq:hubb-ladder:fermionic:mp-eigenvalue-problem}. %
		Indicated are also the two gaps defining the two critical ring\hyp to\hyp center hoppings $s_{\mathrm{c},1}, s_{\mathrm{c},2}$. %
	}%
\end{figure}
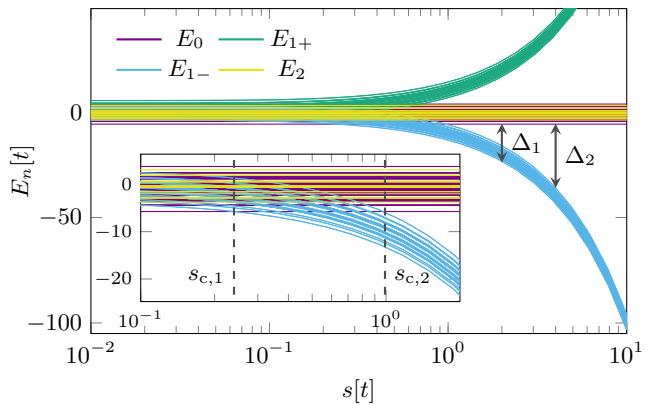%

Using the orthogonality of different Slater determinants, it is a straightforward calculation to find that the general solution of the eigenvalue problem $\hat \Pi_\odot \hat H_{\mathrm{lad}} \hat \Pi_\odot \ket{\mathrm{FS}_N} = E \hat \Pi_\odot \ket{\mathrm{FS}_N}$ reduces to the diagonalization of a $4\times 4$ matrix. %
Fixing a Slater determinant $\ket{\mathrm{FS}_N}$ and two modes $k^\prime, k^\pprime \neq k_0$ so that $\hat c^\nodagger_{k^\prime}\ket{\mathrm{FS}_N} = \ket{\mathrm{FS}_{N-1}}$ as well as $\hat c^\nodagger_{k^\pprime} \hat c^\nodagger_{k^\prime}\ket{\mathrm{FS}_N} = \ket{\mathrm{FS}_{N-2}}$, and labeling the $4$ basis states by their possible occupations of the $k=k_0$ mode $n_{k_0} = 0, 1_\pm, 2$, the resulting eigenvalue problem is of the form%
\begin{equation}%
 	\tikzset{external/export next=false}%
	\tikzsetnextfilename{EW}%
	\begin{tikzpicture}%
		[baseline=(EW.center)]%
		\begin{scope}%
		[%
			node distance = 0.1%
		]%
			\matrix (EW)%
			[%
				matrix of math nodes,%
				left delimiter=(,%
				right delimiter=),%
				nodes in empty cells,%
				nodes = %
				{%
					inner xsep=1pt%
				}%
			]%
			{%
				h_0\phantom{-}	& 					&	 				&					\\%
								& \phantom{h_1-}	&					&					\\%
								& 					&	\phantom{h_1}	&					\\%
								& 					&					&	\phantom{-}h_2	\\%
			};%
		\end{scope}%
		\begin{scope}[on background layer]%
			\node [inner ysep=0pt, inner xsep=1pt, fit=(EW-2-2)(EW-3-3),rounded corners,fill=gray!20,draw=gray!50,fill opacity=0.5] (h1) {};%
			\node at (h1) {$h_1$};%
		\end{scope}%
	\end{tikzpicture}%
	\left(%
		\begin{array}{c}%
			\alpha_0 \\ \alpha_{1_+} \\ \alpha_{1_-} \\ \alpha_2%
		\end{array}%
	\right)%
	=%
	E %
	\left(%
		\begin{array}{c}%
			\alpha_0 \\ \alpha_{1_+} \\ \alpha_{1_-} \\ \alpha_2%
		\end{array}%
	\right)%
	\label{eq:hubb-ladder:fermionic:mp-eigenvalue-problem}%
\end{equation}%
with $h_0 = \braket{0|\hat H_\mathrm{lad}|0}$,  $h_1 = \braket{1_\mu|\hat H_\mathrm{lad}|1_{\mu^\prime}}$ for $\mu,\mu^\prime \in \left\{+,-\right\}$, and $h_2 =  \braket{2|\hat H_\mathrm{lad}|2}$. %
Note the block\hyp diagonal structure that reflects the different $k=k_0$ parities, i.e.,%
\begin{align}%
	\mathrm{e}^{\mathrm{i}\pi\hat n_{k_0}} \ket{n_{k_0}} = %
	\begin{cases}%
		\phantom{-}\ket{n_{k_0}}\, , &\text{if $n_{k_0}=0,2$,} \\%
		-\ket{n_{k_0}}\, , &\text{if $n_{k_0}=1_+,1_-$.}%
	\end{cases}%
\end{align}%
We emphasize the existance of a hidden $\mathbb Z_2$ symmetry of the many\hyp body eigenstates.
This symmetry is an immediate consequence of the modulation of the hopping to the center site, i.e., it characterizes the $k_0$\hyp occupation. %
Furthermore, condensation requires a breaking of particle number conservation on the outer ring, which is possible only in the $n_{k_0}=1_\pm$ subspace. %
Thus, an odd $\mathbb Z_2$ symmetry of the ground state signals the formation of a \gls{BEC}.
Upon solving~\cref{eq:hubb-ladder:fermionic:mp-eigenvalue-problem}, a special structure of the many\hyp body spectrum appears that is characterized by a clustering of eigenstates belonging to the same $k_0$\hyp parity sector, which is exemplified in~\cref{fig:mp-eigenvalues}.
Therefrom, for a given filling fraction $\rho=N/L$ we can extract the scaling of two critical parameters separating the low\hyp lying odd\hyp parity cluster (blue in~\cref{fig:mp-eigenvalues}), which hosts the \gls{BEC} ground state, from the remaining eigenstates.
In what follows we set $t\equiv 1$ as unit of energy. %
The first critical hopping $\tilde s_{\mathrm{c},1}$ and gap $\Delta_{1}$ arise once the clustered odd\hyp parity eigenstates constitute the overall ground state, indicating the condensation of bosons into the $k_0$ mode (abbreviating $X_\rho = \frac{\sin(\pi\rho)}{\pi}$): %
\begin{align}%
	\tilde s_{\mathrm{c},1}%
	&=%
	\sqrt{8X_\rho} \; , \\%
	\Delta_{1}%
	&=%
	-\left(1+2X_\rho \right) + \sqrt{\left(1-2X_\rho \right)^2 + \tilde s^2}%
\end{align}%
The second critical hopping is defined by the complete separation of the odd\hyp parity cluster from the even\hyp parity many\hyp particle eigenstates: %
\begin{align}%
	\tilde s_{\mathrm{c},2}%
	&=%
	4L\sqrt{X_\rho^2 + \mathcal{O}(L^{-1})} \; , \\%
	\Delta_{2}%
	&=%
	-4L X_\rho - \left(1-2X_\rho \right) + \sqrt{\left(1+2X_\rho \right)^2 + \tilde s^2} \; .%
\end{align}%
Note that $\tilde s > \tilde s_{\mathrm{c},2}$ implies that scattering between states with even and odd $k_0$ parity, caused by external perturbations, can only occur if the energy barrier $\Delta_{2}$ can be overcome. %
\paragraph{\label{sec:interactions}Interactions on the outer ring.---}%
\begin{figure}%
	\centering%
	\subfloat[\label{fig:condensate_fraction}]%
	{%
		\ifthenelse{\boolean{buildtikzpics}}%
		{%
			\tikzsetnextfilename{condensate_fraction}%
			\begin{tikzpicture}%
				\begin{axis}%
				[%
					width				= 	0.49\textwidth,%
					height				=	0.25\textheight,%
					xlabel				= 	{$s$},%
					ylabel				= 	{$n_{k_0}/n_\mathrm{max}\vphantom{L}$},%
					legend style		= 	{draw=none, fill=none, draw opacity=0, fill opacity=0},%
					xmode				=	log,
				]%
					\def\Nfacs{0/2, 1/4, 2/8, 3/16}%
					\def\Vs{0/0, 1/0.1, 2/0.5, 3/1.0}%
					\foreach \NfacIt/\Nfac in \Nfacs%
					{%
						\pgfmathsetmacro{\marker}{\gpmarkers[mod(\NfacIt,dim(\gpmarkers))]}%
						\pgfmathsetmacro{\n}{int(32.0/\Nfac)}%
						\edef\markerlegend%
						{%
							\noexpand\addlegendentry{$\rho=\noexpand\nicefrac{1}{\n}$};%
							\noexpand\addlegendimage{mark=\marker, color=black!60, thick};%
							\noexpand\label{pl:marker:\NfacIt}
						}\markerlegend%
					}%
					\node[text width=6em, align=left, anchor=north west] at (axis cs:0.0005,1) 
					{
						\foreach \NfacIt/\Nfac in \Nfacs%
						{%
							\pgfmathsetmacro{\marker}{\gpmarkers[mod(\NfacIt,dim(\gpmarkers))]}%
							\pgfmathsetmacro{\n}{int(32.0/\Nfac)}%
							\edef\markerlegend%
							{%
								\noexpand\ref{pl:marker:\NfacIt} $\rho=\noexpand\nicefrac{1}{\n}$
								\noexpand\\[0.2em]
							}\markerlegend
						}
					};
					\foreach \VIt/\V in \Vs%
					{%
						\pgfmathsetmacro{\cl}{\gpcolors[mod(\VIt,dim(\gpcolors))]}%
						\edef\colorlegend%
						{%
							\noexpand\addlegendentry{$V=\V$};%
							\noexpand\addlegendimage{color=\cl, thick};%
							\noexpand\label{pl:cl:\VIt}
						}\colorlegend%
					}%
					\node[text width=6em, align=left, anchor=south east] at (axis cs:15,0.05) 
					{
						\foreach \VIt/\V in \Vs%
						{%
							\pgfmathsetmacro{\cl}{\gpcolors[mod(\VIt,dim(\gpcolors))]}%
							\edef\colorlegend%
							{%
								\noexpand\ref{pl:cl:\VIt} $V=\V$
								\noexpand\\[0.2em]
							}\colorlegend
						}
					};
					\draw[dashed, thick, black] (axis cs:20,1) -- (axis cs:0.0004,1);%
	
					\foreach \NfacIt/\Nfac in \Nfacs%
					{%
						\pgfmathsetmacro{\marker}{\gpmarkers[mod(\NfacIt,dim(\gpmarkers))]}%
						\foreach \VIt/\V in \Vs%
						{%
							\pgfmathsetmacro{\cl}{\gpcolors[mod(\VIt,dim(\gpcolors))]}%
							\edef\plot%
							{%
								\noexpand\addplot%
								[%
									mark	=	\marker,%
									color	=	\cl,%
									thick,%
									forget plot,%
								]%
									table%
									[%
										x expr	= %
										{%
											\noexpand\thisrowno{1}==\V%
											?%
											(%
												\noexpand\thisrowno{2}==\Nfac%
												?%
												-\noexpand\thisrowno{0}%
												:%
												NaN%
											)%
											:%
											NaN%
										},%
										y expr	= \noexpand\thisrowno{3},%
									]%
								{../data/fitted_fft_k0.dat};%
							}\plot%
						}%
					}%
				\end{axis}%
			\end{tikzpicture}%
		}%
		{%
			\includegraphics{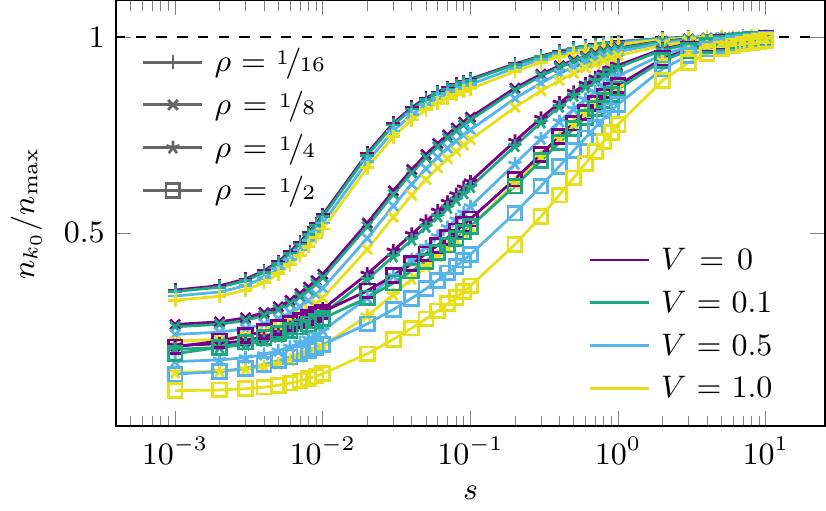}%
		}%
	}%
	\subfloat[\label{fig:condensate_fraction_vs_V_s}]%
	{%
		\ifthenelse{\boolean{buildtikzpics}}%
		{%
			\tikzsetnextfilename{condensate_fraction_vs_V_s}%
			\begin{tikzpicture}%
				\begin{axis}%
				[%
					ymax			=	1.05,%
					xlabel			=	{$\nicefrac{V}{\tilde s}$},%
					ylabel			=	{$n_{k_0}(L)/n_\mathrm{max}(L)$},%
					height			=	0.25\textheight,%
					width			=	0.49\textwidth,%
					xmode			=	log,%
					xmax			=	2,%
				]%
					\draw[dashed, thick, black] (axis cs:0.001,1) -- (axis cs:0.172,1);%
					\addplot%
					[%
						mark	=	+,%
						color	=	colorA,%
						thick,%
					]%
						table%
						[%
							x expr	= %
							{%
								\thisrowno{0}==33%
								?
								(
									\thisrowno{2}==0.5%
									?%
									(%
										round(\thisrowno{3}*32.0/(1.0*\thisrowno{0}-1.0))==1.0*16%
										?%
										\thisrowno{2}/(-\thisrowno{1}*sqrt(\thisrowno{0}))%
										:%
										NaN%
									)%
									:%
									NaN%
								)%
								:%
								NaN%
							},%
							y expr	= \thisrowno{4}/(\thisrowno{3}*(\thisrowno{0}-(\thisrowno{3})+1.0)/(\thisrowno{0})),%
						]%
					{../data/fft_k0_vs_s.dat};%
					\label{pl:cfVs:L33:V0p5:Nfac16}%
					\addplot%
					[%
						mark	=	o,%
						color	=	colorB,%
						thick,%
					]%
						table%
						[%
							x expr	= %
							{%
								\thisrowno{0}==65%
								?
								(
									\thisrowno{2}==1.0%
									?%
									(%
										round(\thisrowno{3}*32.0/(1.0*\thisrowno{0}-1.0))==1.0*16%
										?%
										\thisrowno{2}/(-\thisrowno{1}*sqrt(\thisrowno{0}))%
										:%
										NaN%
									)%
									:%
									NaN%
								)%
								:%
								NaN%
							},%
							y expr	= \thisrowno{4}/(\thisrowno{3}*(\thisrowno{0}-(\thisrowno{3})+1.0)/(\thisrowno{0})),%
						]%
					{../data/fft_k0_vs_s.dat};%
					\label{pl:cfVs:L65:V1:Nfac16}%
					\addplot%
					[%
						mark	=	x,%
						color	=	colorA,%
						thick,%
					]%
						table%
						[%
							x expr	= %
							{%
								\thisrowno{0}==65%
								?
								(
									\thisrowno{2}==0.5%
									?%
									(%
										round(\thisrowno{3}*32.0/(1.0*\thisrowno{0}-1.0))==1.0*16%
										?%
										\thisrowno{2}/(-\thisrowno{1}*sqrt(\thisrowno{0}))%
										:%
										NaN%
									)%
									:%
									NaN%
								)%
								:%
								NaN%
							},%
							y expr	= \thisrowno{4}/(\thisrowno{3}*(\thisrowno{0}-(\thisrowno{3})+1.0)/(\thisrowno{0})),%
						]%
					{../data/fft_k0_vs_s.dat};%
					\label{pl:cfVs:L65:V0p5:Nfac16}%
					\addplot%
					[%
						mark	=	pentagon,%
						color	=	colorB,%
						thick,%
					]%
						table%
						[%
							x expr	= %
							{%
								\thisrowno{0}==129%
								?
								(
									\thisrowno{2}==1.0%
									?%
									(%
										round(\thisrowno{3}*32.0/(1.0*\thisrowno{0}-1.0))==1.0*16%
										?%
										\thisrowno{2}/(-\thisrowno{1}*sqrt(\thisrowno{0}))%
										:%
										NaN%
									)%
									:%
									NaN%
								)%
								:%
								NaN%
							},%
							y expr	= \thisrowno{4}/(\thisrowno{3}*(\thisrowno{0}-(\thisrowno{3})+1.0)/(\thisrowno{0})),%
						]%
					{../data/fft_k0_vs_s.dat};%
					\label{pl:cfVs:L129:V1:Nfac16}%
					\addplot%
					[%
						mark	=	star,%
						color	=	colorA,%
						thick,%
					]%
						table%
						[%
							x expr	= %
							{%
								\thisrowno{0}==129%
								?
								(
									\thisrowno{2}==0.5%
									?%
									(%
										round(\thisrowno{3}*32.0/(1.0*\thisrowno{0}-1.0))==1.0*16%
										?%
										\thisrowno{2}/(-\thisrowno{1}*sqrt(\thisrowno{0}))%
										:%
										NaN%
									)%
									:%
									NaN%
								)%
								:%
								NaN%
							},%
							y expr	= \thisrowno{4}/(\thisrowno{3}*(\thisrowno{0}-(\thisrowno{3})+1.0)/(\thisrowno{0})),%
						]%
					{../data/fft_k0_vs_s.dat};%
					\label{pl:cfVs:L129:V0p5:Nfac16}%
					\addplot%
					[%
						mark	=	square,%
						color	=	colorB,%
						thick,%
					]%
						table%
						[%
							x expr	= %
							{%
								\thisrowno{0}==257%
								?
								(
									\thisrowno{2}==1.0%
									?%
									(%
										round(\thisrowno{3}*32.0/(1.0*\thisrowno{0}-1.0))==1.0*16%
										?%
										\thisrowno{2}/(-\thisrowno{1}*sqrt(\thisrowno{0}))%
										:%
										NaN%
									)%
									:%
									NaN%
								)%
								:%
								NaN%
							},%
							y expr	= \thisrowno{4}/(\thisrowno{3}*(\thisrowno{0}-(\thisrowno{3})+1.0)/(\thisrowno{0})),%
						]%
					{../data/fft_k0_vs_s.dat};%
					\label{pl:cfVs:L257:V1:Nfac16}%
					\node[anchor=south west] at (axis description cs:0.075,0.1) (inset) {};%
					\node%
					[%
						anchor			=	north east%
					] %
						at (axis description cs:1,0.9975)%
						{%
							\setlength{\tabcolsep}{1pt}%
							\scriptsize%
							\begin{tabular}{ccc}%
								\splittedTableHeader{$V$}{$L[\rho]\phantom{mm}$}	&	$\nicefrac12$					&	$1$								\\%
								$33\, [\nicefrac12]$								&	\ref{pl:cfVs:L33:V0p5:Nfac16}  	&									\\%
								$65\, [\nicefrac12]$								&	\ref{pl:cfVs:L65:V0p5:Nfac16}	&	\ref{pl:cfVs:L65:V1:Nfac16}		\\%
								$129\, [\nicefrac12]$								&	\ref{pl:cfVs:L129:V0p5:Nfac16}	&	\ref{pl:cfVs:L129:V1:Nfac16}	\\%
								$257\, [\nicefrac12]$								&									&	\ref{pl:cfVs:L257:V1:Nfac16}	%
							\end{tabular}%
						};%
				\end{axis}%
				\begin{axis}%
				[%
					at					=	(inset.south west),%
					anchor				=	south west,%
					height				=	0.145\textheight,%
					width				=	0.265\textwidth,%
					xticklabel style	=	{font=\tiny},%
					yticklabel style	=	{font=\tiny},%
					xmode				=	log,%
					xmax				=	2,%
				]%
					\draw[dashed, thick, black] (axis cs:0.001,1) -- (axis cs:2,1);%
					\addplot%
					[%
						mark		=	+,%
						mark size	=	1pt,
						color		=	colorC,%
					]%
						table%
						[%
							x expr	= %
							{%
								\thisrowno{0}==33%
								?
								(
									\thisrowno{2}==0.5%
									?%
									(%
										round(\thisrowno{3}*32.0/(1.0*\thisrowno{0}-1.0))==1.0*2%
										?%
										\thisrowno{2}/(-\thisrowno{1}*sqrt(\thisrowno{0}))%
										:%
										NaN%
									)%
									:%
									NaN%
								)%
								:%
								NaN%
							},%
							y expr	= \thisrowno{4}/(\thisrowno{3}*(\thisrowno{0}-(\thisrowno{3})+1.0)/(\thisrowno{0})),%
						]%
					{../data/fft_k0_vs_s.dat};%
					\label{pl:cfVs:L33:V0p5:Nfac2}%
					\addplot%
					[%
						mark		=	o,%
						mark size	=	1pt,
						color		=	colorD,%
					]%
						table%
						[%
							x expr	= %
							{%
								\thisrowno{0}==65%
								?
								(
									\thisrowno{2}==1.0%
									?%
									(%
										round(\thisrowno{3}*32.0/(1.0*\thisrowno{0}-1.0))==1.0*2%
										?%
										\thisrowno{2}/(-\thisrowno{1}*sqrt(\thisrowno{0}))%
										:%
										NaN%
									)%
									:%
									NaN%
								)%
								:%
								NaN%
							},%
							y expr	= \thisrowno{4}/(\thisrowno{3}*(\thisrowno{0}-(\thisrowno{3})+1.0)/(\thisrowno{0})),%
						]%
					{../data/fft_k0_vs_s.dat};%
					\label{pl:cfVs:L65:V1:Nfac2}%
					\addplot%
					[%
						mark		=	x,%
						mark size	=	1pt,
						color		=	colorC,%
					]%
						table%
						[%
							x expr	= %
							{%
								\thisrowno{0}==33%
								?
								(
									\thisrowno{2}==0.5%
									?%
									(%
										round(\thisrowno{3}*32.0/(1.0*\thisrowno{0}-1.0))==1.0*8%
										?%
										\thisrowno{2}/(-\thisrowno{1}*sqrt(\thisrowno{0}))%
										:%
										NaN%
									)%
									:%
									NaN%
								)%
								:%
								NaN%
							},%
							y expr	= \thisrowno{4}/(\thisrowno{3}*(\thisrowno{0}-(\thisrowno{3})+1.0)/(\thisrowno{0})),%
						]%
					{../data/fft_k0_vs_s.dat};%
					\label{pl:cfVs:L33:V0p5:Nfac8}%
					\addplot%
					[%
						mark		=	square,%
						mark size	=	1pt,
						color	=	colorD,%
					]%
						table%
						[%
							x expr	= %
							{%
								\thisrowno{0}==65%
								?
								(
									\thisrowno{2}==1.0%
									?%
									(%
										round(\thisrowno{3}*32.0/(1.0*\thisrowno{0}-1.0))==1.0*8%
										?%
										\thisrowno{2}/(-\thisrowno{1}*sqrt(\thisrowno{0}))%
										:%
										NaN%
									)%
									:%
									NaN%
								)%
								:%
								NaN%
							},%
							y expr	= \thisrowno{4}/(\thisrowno{3}*(\thisrowno{0}-(\thisrowno{3})+1.0)/(\thisrowno{0})),%
						]%
					{../data/fft_k0_vs_s.dat};%
					\label{pl:cfVs:L65:V1:Nfac8}%
					\node%
					[%
						inner sep		=	0pt,%
						anchor			=	south west,%
						scale			=	0.65,%
						text width		=	18em,%
						align			=	left,%
					]%
						at (axis description cs:0.1,0.05)%
						{%
							\setlength{\tabcolsep}{0pt}%
							\tiny%
							\begin{tabular}{ccc}%
								\splittedTableHeader{$V$}{$L[\rho]\phantom{mm}$}	&	$\nicefrac12$					&	$1$								\\%
								$33\, [\nicefrac1{16}]$								&	\ref{pl:cfVs:L33:V0p5:Nfac2}  	&									\\%
								$65\, [\nicefrac1{16}]$								&									&	\ref{pl:cfVs:L65:V1:Nfac2}		\\%
								$33\, [\nicefrac14]$								&	\ref{pl:cfVs:L33:V0p5:Nfac8}  	&									\\%
								$65\, [\nicefrac14]$								&									&	\ref{pl:cfVs:L65:V1:Nfac8}		\\%
							\end{tabular}%
						};%
				\end{axis}%
			\end{tikzpicture}%
		}%
		{%
			\includegraphics{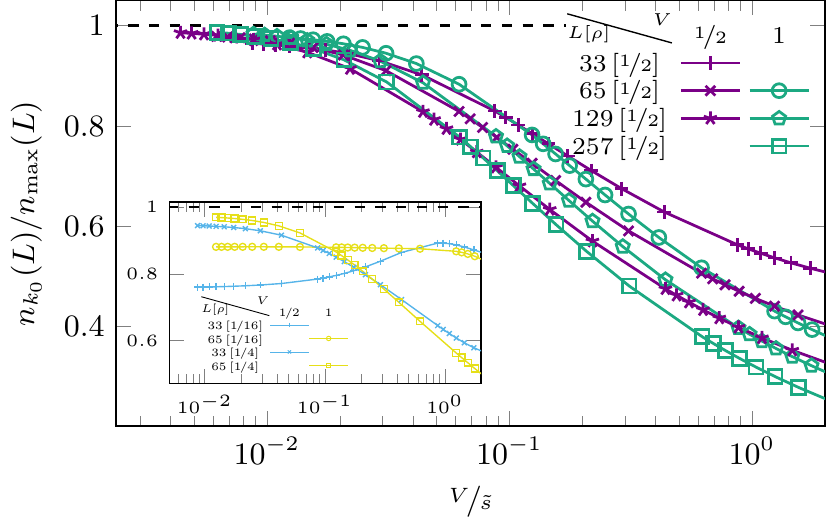}%
		}%
	}%
	\caption%
	{%
		\label{fig:condensate-fraction:normalized}%
		\protect\subref{fig:condensate_fraction} Ground\hyp state \gls{BEC} condensate fraction $n(s)$ normalized to the maximally possible value $n_\mathrm{max}$~\cite{Tennie2017} and extrapolated to the thermodynamic limit. %
		\protect\subref{fig:condensate_fraction_vs_V_s} Asymptotically the normalized condensate fraction for a fixed number of lattice sites is a function of the ratio $\nicefrac{V}{\tilde s}$, only. %
		Results are shown for different \gls{NN} interaction strengths $V$ and densities on the ring $\rho$. %
		Note that for very small fillings $\rho=\nicefrac{1}{16}$ (inset of \protect\subref{fig:condensate_fraction_vs_V_s}), there are significant deviations of the observed connection between the condensate fraction and the ratio $\nicefrac{V}{\tilde s}$. %
		This originitates from the flat single\hyp particle dispersion around $k=\pi$ (see \cref{fig:wheel-to-ladder:sp-dispersion}). %
		Thereby, the complete separation of the odd\hyp parity states (controlled by $\Delta_2$) occurs already for small ring\hyp to\hyp center hoppings, mainly independent on the number of lattice sites. %
	}%
\end{figure}
The analytical solution and, in particular, the property of \gls{BEC} ground states exhibiting odd $k=k_0$ parity allows to draw some striking conclusions on the stability of the \gls{BEC} in the presence of local perturbations on the outer ring. %
Adding interactions acting on a finite subset of outer ring sites only, we note that in general the single\hyp particle description breaks down in favor of a Luttinger liquid~\cite{luttinger,HaldaneJPC,VolkerPRB1992,review_Schulz}. %
Therefore, interactions generically couple the two parity sectors in the $k_0$ subspace and one might expect a breaking of the $\mathbb Z_2$ symmetry. %
However, mixing of the $k_0$ parity sectors caused by local interactions connecting $d$ neighboring sites on the outer ring is of the order of $\sim \frac{Vd}{\tilde s}$ where $V$ is the largest interaction strength. %
The consequence is that increasing the number of lattice sites, i.e., the center site's coordination number, $\mathbb Z_2$ symmetry of the $k_0$ modes is approximately restored, stabilizing \gls{BEC} in the presence of interactions on the outer ring. %
We numerically checked the robustness of the \gls{BEC} in the thermodynamic limit for $k_0=0$ and finite values of the ring\hyp to\hyp center hopping. %
To this end, we calculated the ground\hyp state occupation $n_{k_0}(s,L)$ of the $k_0=0$ mode~\cite{supersolid_penrose,Yang1962} using \gls{DMRG}. %
Normalizing with respect to the upper bound on the condensate occupation $n_\mathrm{max}(L)=L \rho \left(1-\rho + 1/L\right)$~\cite{Tennie2017}, we extrapolated the condensate fraction into the thermodynamic limit $n_{k_0}(s)/n_\mathrm{max} = \lim\limits_{L\rightarrow\infty} \frac{n_{k_0}(s,L)}{n_\mathrm{max}(L)}$ (see~\cite{suppMat} for the details). %
The resulting extrapolations are shown in~\cref{fig:condensate_fraction} for interaction strengths between $V=0$ and $V=1$ and particle densities between $\rho=\nicefrac{1}{16}$ and $\nicefrac{1}{2}$. %
Note that even though there is a renormalization of the overall condensate fraction, we always observe a finite condensate density in the thermodynamic limit, even for strong interactions and high particle densities. %
To further demonstrate the asymptotic robustness of the $\mathbb Z_2$ symmetry protection, in~\cref{fig:condensate_fraction_vs_V_s} the dependency of the condensate fraction at finite system sizes and as a function of $\nicefrac{V}{\tilde s}$ is shown. %
We also observe the behavior expected from our previous analysis, namely that the condensate occupation dominantly depends on the system size and the ratio between the interaction and the extensively scaling renormalized ring\hyp to\hyp center hopping $\tilde s = s \sqrt L$. %
In accordance to the scaling of $\Delta_2$, the maximally possible condensation is reached if $\tilde s \sim L$. %
We emphasize that these relations can be translated to experimental realizations to determine the necessary coordination number, i.e., number of sites on the ring, to detect \gls{BEC} in the presence of interactions. %
\paragraph{\label{sec:conclusion}Conclusion.---}%
We introduced a solution strategy for models on a wheel with $k_0$ modulated ring\hyp to\hyp center hopping $s_j = s\mathrm{e}^{\mathrm{i}k_0j}$, which we applied to a system of \gls{HCB}~\cite{Dongen1991,Vidal2011,Tennie2017,Mate2021}. %
Our central finding is the protection of a \gls{BEC} by a $\mathbb Z_2$ symmetry emerging from the model\hyp specific modulation of the hopping to the center site. %
We traced back this remarkable feature to an extensively scaling separation $\propto s\sqrt L$ of the $k=k_0$ single\hyp particle modes, generated from the extensive coordination of the center with the ring sites. %
This scaling renders the \gls{BEC} robust against local perturbations on the ring. %
We demonstrated this feature numerically by calculating the \gls{HCB} $k_0$\hyp condensate fraction of the ground state in the presence of \gls{NN} interactions and for various particle number densities. %
Our calculations clearly show the protection of \gls{BEC} where the condensate fraction is controlled by the ratio $\frac{V}{s \sqrt L}$ and approaches the maximally possible value~\cite{Tennie2017}, even in the presence of strong interactions. %

Our findings imply important consequences for both experimental and theoretical realizations of wheel geometries in general. %
First of all, a particular single\hyp particle mode can be gapped out by a proper modulation of the ring\hyp to\hyp center hopping, allowing the general protection of ordered phases that are characterized by a certain wave vector.
We believe that such a modulation of the ring\hyp to\hyp center hopping provides an experimentally feasible approach to realize exotic, finite\hyp momentum \gls{BEC} in the framework of ultracold or Rydberg atoms~\cite{Lim2008,Hick2010,DiLiberto2011,Lin2011,Summy2016,Mangaonkar2020,Eiles2021}.
Second, there is a many\hyp body gap separating the \gls{BEC}\hyp carrying states from the remaining spectrum $\sim s \sqrt L$, i.e., large gaps can be realized by increasing the coordination number of the center site. %
The resulting robustness against interactions on the ring can be exploited to increase critical temperatures for phase transitions into otherwise highly fragile quantum phases. %
Possible applications are mesoscopic setups where a conducting center site may be contacted to one\hyp dimensional ring geometries via tunnel contacts, allowing the stabilization of ordered states on the ring against perturbations. %
Such a scaling could also be exploited to increase the stability of superconducting qubits by means of an all\hyp to\hyp all\hyp coupling of a set of noisy stabilizer qubits to a central qubit~\cite{Krantz2019}.
Moreover, we believe that the wheel\hyp to\hyp ladder mapping could prove useful in the analysis of hidden fermions~\cite{Sakai2016}.
Further interesting questions are the incorporation of disorder on both the ring and center site, as well as the effect of (artificial) gauge fields and a rescaled ring\hyp to\hyp center hopping $s \rightarrow \frac{s}{\sqrt{L}}$ with regard to the crossover from one to an infinite number of dimensions. %
\paragraph*{\label{sec:acknowledgement}Acknowledgment.---}%
We thank F.~Grusdt and U.~Schollw\"ock for carefully reading the manuscript. %
Furthermore, we thank M.~Bramberger and M.~Grundner for very fruitful discussions. %
TK acknowledges financial support by the ERC Starting Grant from the European Union's Horizon 2020 research and innovation program under grant agreement number 758935. %
FAP acknowledges funding by the Deutsche Forschungsgemeinschaft (DFG, German Research Foundation) via Research Unit FOR 2414 under project number 277974659. %
RHW, FAP and SP acknowledge support from the Munich Center for Quantum Science and Technology. %
The authors gratefully acknowledge the funding of this project by computing time provided by the Paderborn Center for Parallel Computing (PC$^2$).
\bibliography{Literatur}%
\pagebreak
\widetext
\begin{center}
\textbf{\large Supplemental Materials}
\end{center}
\setcounter{equation}{0}%
\setcounter{figure}{0}%
\setcounter{table}{0}%
\setcounter{page}{1}%
\makeatletter%
\renewcommand{\theequation}{S\arabic{equation}}%
\renewcommand{\thefigure}{S\arabic{figure}}%
\renewcommand{\bibnumfmt}[1]{[S#1]}%
\makeatother
%
%
\section{\label{sec:supp:wheel-to-ladder}Wheel\hyp to\hyp Ladder mapping}
Here, we outline the mapping from the wheel geometry to the projected ladder in detail.
The wheel Hamiltonian is given by
\begin{align}
	\hat H
	& =
	- t \sum_{j=0}^{L-1} \left( \hat h^\dagger_j \hat h^\nodagger_{j+1} + \hat h^\dagger_{j+1} \hat h^\nodagger_j \right)
	-  \sum_{j=0}^{L-1} \left(  s_j \hat h^\dagger_j \hat h^\nodagger_\odot + \mathrm{h.c.} \right) \;, \label{eq:supp:hamiltonian:wheel}
\end{align}
with $\hat h^{(\dagger)}_j\left( \hat h^{(\dagger)}_\odot\right) $ describing hardcore bosonic degrees of freedom located on the outer ring (center site) and $s_j = \mathrm{e}^{\mathrm{i} k_0 j} $ with $k_0 = \frac{2\pi}{L} n$ ($n \in \mathbb Z$) a reciprocal lattice vector.
We consider periodic boundary conditions on the outer ring, i.e., $\hat h^{(\dagger)}_L = \hat h^{(\dagger)}_0$.
The model is defined on the tensor product Hilbert space $\mathcal H_{L+1} = \mathcal H^{\otimes L}_{2} \otimes \mathcal H_{\odot}$, where $\mathcal H_{2} \left( \mathcal H_{\odot}\right) $ is the single\hyp particle Hilbert space of a hardcore boson on the outer ring (center site).
We define an enlarged Hilbert space of two concentric   rings, where the Hilbert space of the inner ring is a copy of the Hilbert space of the outer ring. 
\begin{equation}
\begin{aligned}
	\mathcal{H}_{L+1} = \mathcal H_L \otimes &\mathcal H_\odot  = \mathcal H^{\otimes L}_2 \otimes \mathcal H_2\\
	&\downarrow\\
	\mathcal H_{\mathrm{lad},L} = \mathcal H_L \otimes&\mathcal H_{\odot, L} = \mathcal H^{\otimes L}_2 \otimes \mathcal H^{\otimes L}_2
	\notag
\end{aligned}
\end{equation}
To describe hopping to the inner ring, we introduce $\hat h^{(\dagger)}_{\odot,j}$ as the corresponding ladder operators acting on the sites of the inner ring.
Note that sites with the same index on the inner and outer ring can be aligned vertically, yielding a ladder geometry.
We introduce the second quantization basis for the inner ring
\begin{equation}
	\ket{n_{\odot, 1}, n_{\odot, 2},\ldots, n_{\odot, L}}_\odot \in \mathcal H_{\odot, L}\notag
\end{equation}
as well as the vacuum state $\ket{\varnothing}_\odot \in \mathcal H_{\odot, L}$.
In order to represent the same physical situation as the wheel system, the total occupation of all sites on the inner ring must be either zero or one.
Furthermore, we enforce the allowed states in $\mathcal H_{\odot, L}$ to transform under rotations of the inner ring, which is an a priori constraint so far, but will turn out to be very useful in the forthcoming discussion.
We introduce $\mathcal{\tilde{H}}_{\odot}$ by the set of states $\ket{\omega}_\odot$ that meet these constraints
\begin{equation}
	\tilde{\mathcal H}_\odot = \operatorname{span} \left\{ \ket{\omega}_\odot \in \mathcal H_{\odot, L}\right\}.
	\notag
\end{equation}
On this subspace of $\mathcal{H}_{\odot,L}$, we then must have
\begin{equation}
\begin{aligned}
	\label{eq:supp:constraints}
		\fvN_{\odot}\hspace{-.15em} \braket{\omega| \hat N_\odot| \omega }_\odot
		& = 
		\sum_j  \fv_{\odot}\hspace{-.15em} \braket{\omega| n_{\odot,j}| \omega }_\odot = 0,1\\
	\fvR_{\odot}\hspace{-.15em} \braket{\omega| [\hat R_{\odot}]^n | \omega }_\odot
	&= 
	\mathrm{e}^{\mathrm i q} \fvR_{\odot}\hspace{-.15em} \braket{\omega| [\hat R_{\odot}]^{n-1} | \omega }_\odot \; , \quad q = \frac{2\pi}{L} m, \; m \in \mathbb Z
\end{aligned}
\end{equation}
where $n = 0,1,\ldots,L-1$ and $\hat R_\odot$ applies a rotation to the inner ring:
\begin{align}
	\hat R_\odot: 
	\mathcal H_{\odot,L} &\longrightarrow \mathcal H_{\odot, L} \notag \\
	\ket{n_{\odot, 1}, \ldots, n_{\odot, L}}_\odot & \longmapsto \ket{n_{\odot,2}, \ldots, n_{\odot,1}}_\odot \; ,\notag
\end{align}
and addition in the site indices is performed $\operatorname{modulo} L$.
The allowed states satisfying the above constraints are given by
\begin{equation}
	\ket{\omega=0}_\odot
	=
	\ket{\varnothing}_\odot
	\quad \text{and} \quad
	\ket{\omega=1}_\odot
	=
	\frac{1}{\sqrt{L}} \sum_n \mathrm{e}^{\mathrm{i} q n}  [\hat R_\odot]^n \hat h^\dagger_{\odot,j} \ket{\varnothing}_\odot \; ,
	\label{eq:supp:allowed-states}
\end{equation}
for any $j \in \left\{ 1, \ldots, L \right\}$.

Let us from now on identify $\ket{0}_\odot \left( \ket{1}_\odot\right) $ with the empty (occupied) inner site of the wheel.
Having introduced the allowed states, we construct a Hamiltonian exhibiting the same matrix elements in $\mathcal{P} = \mathcal H_L \otimes \tilde{\mathcal H}_\odot$ as~\cref{eq:supp:hamiltonian:wheel} in $\mathcal H_{L+1} $:
\begin{align}
	\hat L_\odot 
	&\equiv
	\underbrace{-t \sum_{j=0}^{L-1} \left( \hat h^\dagger_j \hat h^\nodagger_{j+1} + \mathrm{h.c.} \right)}_{\hat L_{\odot,t}}
	- \sum_{j=0}^{L-1} \left( s_j \hat h^\dagger_j \hat \omega^\nodagger_{\odot} + \mathrm{h.c.}  \right) \; .
		\notag
\end{align}
Here, we defined operators $\hat \omega^\nodagger_{\odot} = \ket{0}_\odot\bra{1}_\odot$ and $\hat \omega^\dagger_\odot =  \ket{1}_\odot\bra{0}_\odot$ via their action on $\ket{1}_\odot$ and $\ket{0}_\odot$, respectively.
In order to obtain a representation of $\hat L_\odot$ in $\mathcal{P}$ we write 
\begin{align}
\hat \omega^{(\dagger)}_{\odot}\ket{\varnothing} = \frac{1}{\sqrt{L}}\sum_l \mathrm{e}^{-\mathrm{i} ql}  \hat h^{(\dagger)}_{\odot,l}\ket{\varnothing} \; ,
\end{align}

where $\hat h^{(\dagger)}_{\odot,l}$ acts on the $l$-th site of the inner ring, as well as a projector that projects to $\mathcal{F}_{\odot,1} \subset \mathcal H_{\odot, L}$, i.e., the Fock space spanned by empty and singly\hyp occupied states on the inner ring:

\begin{align}
	\hat P^\nodagger_\odot
	&=
	\prod_j \hat n^e_{\odot, j}
	+
	\sum_j \hat n_{\odot, j} \prod_{k \neq j} \hat n^e_{\odot, k} \; ,
	\label{eq:projector_P}
\end{align}
with $\hat n^e_{\odot, j} = \hat 1_{\odot, j} - \hat n_{\odot, j}$ and $\hat n_{\odot, j} = \hat h^\dagger_{\odot, j}\hat h^\nodagger_{\odot, j}$.
Since $\hat P^\dagger_\odot = \hat P^\nodagger_\odot$ and $\left(\hat P^\nodagger_\odot\right)^2 = \hat P^\nodagger_\odot$, $\hat P^\nodagger_\odot$ indeed is a projector.
Rewriting $\hat \omega_\odot$ and projecting down to $\mathcal{F}_{\odot,1}$, we obtain
\begin{align}
\hat H_{\mathrm{proj}} = 
\hat P^\nodagger_\odot\hat L_{\odot}\hat P^\nodagger_\odot
	&=
	 \hat P^\nodagger_\odot\hat L_{\odot,t}\hat P^\nodagger_\odot + \frac{s}{\sqrt L} \sum_{j,l} \hat P^\nodagger_\odot\left( \mathrm{e}^{\mathrm{i} k_0 j}\mathrm{e}^{-\mathrm{i} ql} \hat h^\dagger_j \hat h^\nodagger_{\odot,l} + \mathrm{h.c.}\right) \hat P^\nodagger_\odot \notag \\
	 &=
	  \hat P^\nodagger_\odot\hat L_{\odot,t}\hat P^\nodagger_\odot + \frac{s}{\sqrt L} \sum_{l} \mathrm{e}^{-\mathrm{i} ql} \hat P^\nodagger_\odot \left[ \hat R^\dagger_\odot\right]^l \left( \sum_{j} \mathrm{e}^{\mathrm{i} k_0 j} \hat h^\dagger_j \hat h^\nodagger_{\odot,j}\right)\left[ \hat R^\nodagger_\odot\right]^l \hat P^\nodagger_\odot  + \mathrm{h.c.} \notag \\
	 &=
	 \hat P^\nodagger_\odot\hat L_{\odot,t}\hat P^\nodagger_\odot + \frac{s}{\sqrt L} \sum_{l} \sum_{k,k^\prime}
	  \ket{k}_\odot\braket{k|\mathrm{e}^{-\mathrm{i}\left( q+k-k^\prime\right)l} \hat P^\nodagger_\odot \left( \sum_{j} \mathrm{e}^{\mathrm{i} k_0 j} \hat h^\dagger_j \hat h^\nodagger_{\odot,j}\right) \hat P^\nodagger_\odot |k^\prime} {\fv\vspace{.5em}}_{\odot}\hspace{-.15em}\bra{k^\prime} + \mathrm{h.c.} \notag \\
	  &=
	  \hat P^\nodagger_\odot\hat L_{\odot,t}\hat P^\nodagger_\odot + s\sqrt L \sum_{k}
	  \ket{k}_\odot\braket{k| \hat P^\nodagger_\odot \left( \sum_{j} \mathrm{e}^{\mathrm{i} k_0 j} \hat h^\dagger_j \hat h^\nodagger_{\odot,j}\right) \hat P^\nodagger_\odot |k + q}{\fv\vspace{.5em}}_{\odot}\hspace{-.15em}\bra{k + q} + \mathrm{h.c.} \; .
	\label{eq:supp:projected-ladder-rot}	
\end{align}
which now acts on $\mathcal H_L \otimes \mathcal F_{\odot,1}$.
We introduced $\ket{k}_\odot$, the eigenstates of $\hat R_\odot$, labeled by the respective rotation angle $k = \frac{2\pi}{L} n$ with $n\in \left\{0,\ldots,L-1 \right\}$ so that $\hat R_\odot \ket{k}_\odot = e^{\mathrm{i} k} \ket{k}_\odot$.
In the single\hyp particle subspace on the inner ring, these states are given by
\begin{align}
	\label{eq:rot-eigenstates}
	\ket{k}_\odot = \frac{1}{\sqrt{L}} \sum_{j=0}^{L-1} \mathrm{e}^{\mathrm{i} kj} \hat h^\dagger_{\odot,j} \ket{\varnothing}_\odot \; .
\end{align}
In order to obtain a representation that is block\hyp diagonal in the $\ket{k}_\odot$\hyp basis we choose $q=0$:
\begin{align}
	\hat H_{\mathrm{proj}} =  \hat P^\nodagger_\odot \hat H_{\mathrm{lad}} \hat P^\nodagger_\odot\; , \quad
	\hat H_{\mathrm{lad}}
	=
	-t \sum_{j}\left( \hat h^\dagger_j \hat h^\nodagger_{j+1} + \mathrm{h.c.} \right) 
	-  \tilde s\sum_{j} \left(\mathrm{e}^{\mathrm{i} k_0 j} \hat h^\dagger_j \hat h^\nodagger_{\odot, j} + \mathrm{h.c.} \right)\; .
\end{align}
We note that $\hat H_{\mathrm{proj}}$ has long\hyp ranged hoppings while $\hat H_{\mathrm{lad}}$ does not.
Importantly, the eigenstates of $\hat R_\odot$ are eigenstates of $\hat H_{\mathrm{proj}}$ due to $[\hat H_{\mathrm{proj}}, \hat R_\odot] = 0$, while this is not the case for $\hat H_{\mathrm{lad}}$, since $[\hat R_\odot, \hat H_{\mathrm{lad}}] \neq 0$.
Since $[\hat N_\odot, \hat R_\odot] = 0$ holds, with $\hat N_\odot$ being the total particle number on the inner ring, we may set up the simultaneous eigenstates and group them by their corresponding eigenvalues of $\hat R_\odot$:
\begin{align}
	N_\odot &= 0 \Rightarrow \ket{N_\odot = 0, k = 0}_\odot \; \text{is uniquely specified in the subspace with $\hat N_\odot = 0$,} \\
	N_\odot &= 1 \Rightarrow \ket{N_\odot = 1, k = \frac{2\pi}{L} n}_\odot \; \text{are $L$ different states in the subspace $\hat N_\odot = 1$.}
\end{align}
They span the Fock space $\mathcal{F}_{\odot,1}$.
The allowed states~\cref{eq:supp:allowed-states} are obtained by taking only the states with $k = 0$ in 	\cref{eq:rot-eigenstates}.
For brevity, we define  $\ket{\varnothing}_\odot = \ket{N_\odot = 0,  k = 0}_\odot$ and $\ket{k = 0}_\odot = \ket{N_\odot = 1, k = 0}_\odot$
Now, let $\hat \Pi_\odot = \ket{\varnothing}_\odot\bra{\varnothing} + \ket{k = 0}_\odot\bra{k = 0}$ be the projector into the $k = 0$ subspace on the inner ring.
We can thus write the initial Hamiltonian~\cref{eq:supp:hamiltonian:wheel}, mapped to the ladder geometry as
\begin{align}
	\label{eq:supp:hubb-wheel:proj-condition}
	\hat H \equiv \hat \Pi_\odot \hat H_{\mathrm{proj}} \hat \Pi_\odot &= \hat \Pi_\odot \hat P^\nodagger_\odot \hat H_{\mathrm{lad}} \hat P^\nodagger_\odot \hat \Pi_\odot = \hat \Pi_\odot \hat H_{\mathrm{lad}} \hat \Pi_\odot \; .
\end{align}
This establishes the mapping between the eigenstates of $\hat \Pi_\odot \hat H_{\mathrm{lad}} \hat \Pi_\odot$, which will be computed explicitly later on, and the desired eigenstates of~\cref{eq:supp:hamiltonian:wheel}, i.e., we need to find the many\hyp particle eigenstates of~\cref{eq:supp:projected-ladder-rot} projected into the $k = 0$ sector.
For that purpose, we employ a Jordan\hyp Wigner transformation expressing the hardcore bosonic ladder operators in terms of fermionic ones.
Implementing the ladder geometry via the mappings $\hat h^{(\dagger)}_j \mapsto \hat h^{(\dagger)}_{2j}$ and $\hat h^{(\dagger)}_{\odot, j} \mapsto \hat h^{(\dagger)}_{2j+1}$ yields the fermionic ladder operators
\begin{align}
	\hat c^{(\dagger)}_j = \prod_{k<j}\mathrm{e}^{\mathrm{i}\pi \hat n_k} \hat h^{(\dagger)}_{j}\; \text{, with $j=0,\ldots,2L-1$.}
\end{align}
After transforming the Hamiltonian, we reintroduce fermionic operators on the inner and outer ring via $\hat c^{(\dagger)}_{2j} \mapsto \hat c^{(\dagger)}_j$ and $\hat c^{(\dagger)}_{2j+1} \mapsto \hat c^{(\dagger)}_{\odot, j}$ and obtain
\begin{align}
	\hat H_{\mathrm{lad}}
	&=
	t \sum_j \left( \hat c^\dagger_j \mathrm{e}^{\mathrm{i}\pi \hat n_{\odot,j}}\hat c^\nodagger_{j+1} + \mathrm{h.c.} \right)
	-
\tilde s \sum_j \left(	\mathrm{e}^{\mathrm{i}k_0 j} \hat c^\dagger_j \hat c^\nodagger_{\odot,j} + \mathrm{h.c.} \right) \;.
\end{align}
\section{\label{sec:supp:sp-solution}Solution of the Single\hyp Particle Problem}
In the single\hyp particle subspace, the Jordan\hyp Wigner transformed ladder Hamiltonian becomes 
\begin{align}
	\hat H_{\mathrm{lad}}
	&=
	t \sum_j \left( \hat c^\dagger_j \hat c^\nodagger_{j+1} + \mathrm{h.c.} \right)
	-
	\tilde s \sum_j \left(	\mathrm{e}^{\mathrm{i}k_0 j} \hat c^\dagger_j \hat c^\nodagger_{\odot,j} + \mathrm{h.c.} \right) \; .
\end{align}
It is instructive to explicitly construct the single\hyp particle eigenstates subject to the projection into the $q=0$ subspace, where for brevity we introduce the decomposition $\hat H_{\mathrm{lad}} = \hat H_{\mathrm{lad}, t} + \hat H_{\mathrm{lad}, s}$.
We begin by considering the action of the summands on rotational eigenstates $\ket{k}_{(\odot)}$ on the outer (inner) ring:
\begin{align}
	\hat H_{\mathrm{lad}, t}\ket{k}\ket{\varnothing}_\odot 
	&= \ket{k}\ket{\varnothing}_\odot
	\quad \text{and} \quad
	\hat H_{\mathrm{lad}, t}\ket{\varnothing}\ket{k}_\odot 
	= 0 \; , \\
	\hat H_{\mathrm{lad}, s}\ket{k}\ket{\varnothing}_\odot 
	&=\sum_{j,l} \mathrm{e}^{-\mathrm{i} k_0 j} \hat c^\dagger_{\odot, j} \hat c^\nodagger_j \frac{ \mathrm{e}^{\mathrm{i} k l}}{\sqrt{L}} \hat c^\dagger_l \ket{\varnothing} \ket{\varnothing}_\odot
	= \ket{\varnothing}\ket{k-k_0}_\odot \; , \\
	\hat H_{\mathrm{lad}, s}\ket{\varnothing} \ket{k}_\odot 
	&= \sum_{j,l} \mathrm{e}^{\mathrm{i} k_0 j} \hat c^\dagger_{j} \hat c^\nodagger_{\odot,j} \frac{ \mathrm{e}^{\mathrm{i} k l}}{\sqrt{L}} \hat c^\dagger_{\odot,l} \ket{\varnothing} \ket{\varnothing}_\odot
	= \ket{k+k_0}\ket{\varnothing}_\odot \; .
\end{align}
For the projected eigenvalue equation this motivates the Ansatz $\hat \Pi_\odot \ket{\psi_k} = \ket{k}\ket{\varnothing}_\odot$,
\begin{align}
	\hat \Pi_\odot \hat H_{\mathrm{lad}}\hat \Pi_\odot \ket{\psi_k} 
	&= 	\hat \Pi_\odot\left( \epsilon_k \ket{k}\ket{\varnothing}_\odot - \tilde s \ket{\varnothing}\ket{k-k_0}_\odot  \right) \\
	&= \epsilon_k \ket{k}\ket{\varnothing}_\odot - \tilde s \hat \Pi_\odot \ket{\varnothing}\ket{k-k_0}_\odot \;,
\end{align}
where $\epsilon_k = 2t\cos(k)$ is the single\hyp particle dispersion relation of non\hyp interacting, spinless fermions.
For $k\neq k_0$, $\hat\Pi_\odot \ket{\psi_k}$ indeed satisfies the eigenvalue equation. In order to determine the $k=k_0$  single\hyp particle eigenstates we make the ansatz $	\hat \Pi_\odot \ket{\psi_{k_0}} = \sum_{k^\prime} c_{k^\prime} \ket{k^\prime}\ket{\varnothing}_\odot + \Delta \ket{\varnothing}\ket{0}_\odot$
\begin{align}
	\hat \Pi_\odot \hat H_{\mathrm{lad}}\hat \Pi_\odot \ket{\psi_k} 
	&=\left(  \sum_{k^\prime} \epsilon_{k^\prime} c_{k^\prime} \ket{k^\prime} - \Delta \tilde s \ket{k_0}\right)\ket{\varnothing}_\odot  - \tilde s c_{k_0} \ket{\varnothing}\ket{0}_\odot \notag \\
	&\overset{!}{=} \epsilon_{k_0} \left( \sum_{k^\prime} c_{k^\prime} \ket{k^\prime}\ket{\varnothing}_\odot + \Delta \ket{\varnothing}\ket{0}_\odot \right)  
\end{align}
Setting $c_{k^\prime} = \delta_{k^\prime,k_0} c_{k_0}$ and choosing $c_{k_0} = 1$, we can solve for $\Delta$
\begin{align}
	&\left( \epsilon_{k_0} - \Delta \tilde s \right) \ket{k_0}\ket{\varnothing}_\odot - \tilde s \ket{\varnothing}\ket{0}_\odot  
	= \epsilon_{k_0} \left( \ket{k_0}\ket{\varnothing}_\odot + \Delta \ket{\varnothing}\ket{0}_\odot \right)  \notag \\
	&\rightarrow \Delta_{\pm} = \frac{\epsilon_{k_0}}{2 \tilde s } \pm \sqrt{\frac{\epsilon_{k_0}^2}{4 \tilde s^2 } + 1} \;,
\end{align}
Thus, for $k=k_0$, there are two orthogonal single\hyp particle eigenstates
\begin{align}
	\ket{\psi_{k_0,\pm}}
	&=
	\psi_{k_0,\pm} \left(	\hat c^\dagger_{k_0} + \Delta_{\pm}\hat c^\dagger_{\odot, k=0}\right) \ket{\varnothing} \;.
	\label{eq:supp:sf-hubb:sp-eigenstates}
\end{align}
In this representation, projecting down the eigenstates into the zero\hyp momentum sector on the inner ring is particularly easy:
\begin{align}
	\label{eq:supp:sf-hubb:proj-eigenstates}
	\ket{k,\pm}
	= 
	\hat \Pi_\odot \ket{\psi_{k,\pm}}
	=
	\begin{cases}
		\psi_{k} \hat c^\dagger_{k} \ket{\varnothing} &\text{if $k \neq k_0$,} \\
		 \psi^\dagger_{\pm} \ket{\varnothing}  \equiv \psi_{\pm} \left(\hat c^\dagger_{k} + \Delta_{\pm}\hat c^\dagger_{\odot, k=0}\right) \ket{\varnothing}  &\text{if $k = k_0$.}
	\end{cases}
\end{align}

Using~\cref{eq:supp:hubb-wheel:proj-condition}, the single\hyp particle problem of the wheel Hamiltonian~\cref{eq:supp:hamiltonian:wheel} is thus solved by expanding the ladder problem in terms of the $\ket{\psi_{k,\mu}}$ and projecting into the $q=0$ subspace
\begin{align}
	\hat H
	&\equiv
	\hat \Pi_\odot \sum_{k,k^\prime}\sum_{\mu,\mu^\prime=\pm} \ket{\psi_{k,\mu}} \braket{\psi_{k,\mu}|\hat \Pi_\odot \hat H_\mathrm{lad} \hat \Pi_\odot| \psi_{k^\prime,\mu^\prime}}\bra{\psi_{k^\prime, \mu^\prime}}\hat \Pi_\odot \notag \\
	&=
	\sum_{k\neq k_0} \epsilon_k \ket{k}\bra{k} \otimes \ket{\varnothing}_\odot\bra{\varnothing} + \sum_{\mu = \pm} \varepsilon_\mu \ket{\psi_{k=k_0,\mu}}\bra{\psi_{k=k_0,\mu}} \; ,
\end{align}
where $\varepsilon_\pm = \frac{1}{2}\left(\varepsilon_0 \pm \operatorname{sgn}(\tilde s)\sqrt{\varepsilon^2_0 + 4 \tilde s^2} \right)$.
For later convenience, we also introduce ladder operators $\hat \psi^{(\dagger)}_{k,\mu}$  annihilating (creating) single\hyp particle modes $\ket{\psi_{k,\mu}}$.
From~\cref{eq:supp:sf-hubb:sp-eigenstates} it can be readily checked that they obey fermionic anticommutation relations
\begin{align}
	\left\{ \hat \psi^{(\dagger)}_{k,\mu}, \hat \psi^{(\dagger)}_{k^\prime,\mu^\prime} \right\} = 0 \quad \text{and} \quad
	\left\{ \hat \psi^{\nodagger}_{k,\mu}, \hat \psi^{\dagger}_{k^\prime,\mu^\prime} \right\} = \delta_{k,k^\prime} \delta_{\mu, \mu^\prime} \; .
\end{align}

\section{\label{sec:supp:mp-solution}Solution of the Many\hyp Particle Problem}
The solution of the full many\hyp particle problem~\cref{eq:supp:hamiltonian:wheel} is done with the help of Slater determinants constructed from the single\hyp particle eigenstates~\cref{eq:supp:sf-hubb:proj-eigenstates}.
Here, the important point is that all many\hyp particle Slater determinants with either empty or doubly occupied $k=k_0$ modes are already eigenstates of the projected ladder~\cref{eq:supp:hubb-wheel:proj-condition} and thereby of the wheel Hamiltonian.
In order to prove this observation, we define for a given set of $N$ modes $\mathbf k_N = \left(k_1, \ldots, k_N \right)$ with $k_l = \frac{2\pi}{L}n_l \neq k_0$ and $0 \leq n_l < L$ projected Slater determinants of single\hyp particle eigenstates of $\hat H_\mathrm{lad}$
\begin{align}
	\ket{\mathbf k_N} = \ket{k_1, \ldots, k_N} = \hat \Pi_\odot \ket{\psi_{k_1}, \cdots, \psi_{k_N}} = \hat c^\dagger_{k_1} \cdots \hat c^\dagger_{k_N} \ket{\varnothing} \; ,
\end{align}
wherein we fixed the global phase by normal ordering the modes: $k_1 < k_2 < \cdots < k_N$.
Within this ordering, Slater determinants with modes $\ket{\psi_{k=k_0,\pm}}$ occupied are always moved to the left and denoted by:
\begin{align}
	\ket{n_+, n_-} \ket{\mathbf k_N} = \left[\hat \psi^{\dagger}_{k_0,+} \right]^{n_+} \left[\hat \psi^{\dagger}_{k_0,-} \right]^{n_-} \ket{\mathbf k_N} \; .
\end{align}
Note that in the main text, we used a more condensed notation for the $\ket{\psi_{k=k_0,\pm}}$ modes, labeling only the overall occupation via the abbreviations $\ket{0,0} \rightarrow \ket{0}$, $\ket{1,0} \rightarrow \ket{1_+}$, $\ket{0,1} \rightarrow \ket{1_-}$ and $\ket{1,1} \rightarrow \ket{2}$.
However, for clarity reasons, here we maintain the extended representation.
Having setup this notation, we consider the action of the Jordan\hyp Wigner\hyp transformed wheel Hamiltonian in the projected ladder presentation on these Slater determinants
\begin{align}
	\hat H \ket{n_+,n_-} \ket{\mathbf k_N}
	&=
	\hat \Pi_\odot \hat H_\mathrm{lad} \hat \Pi_\odot \ket{n_+,n_-} \ket{\mathbf k_N} \notag \\
	&=
	\hat \Pi_\odot \left(
		t\sum_{j=0}^{L-1} \left( 
			\hat c^\dagger_j \mathrm{e}^{\mathrm{i}\pi \hat n_{\odot, j}} \hat c^\nodagger_{j+1} + \mathrm{h.c.}
		\right)
		-
	 \sum_{j=0}^{L-1}\left(
				\tilde s^\nodagger_j\hat c^\dagger_j \hat c^\nodagger_{\odot, j} + \mathrm{h.c.}
		\right)
	\right) \hat \Pi_\odot \ket{n_+,n_-} \ket{\mathbf k_N} \; .
\end{align}
In order to proceed, in the first sum we expand the operators acting on the inner ring in terms of rotations, i.e., perform a Fourier transformation (using $k_n = \frac{2\pi}{L}n$) and apply the projection to the $q=0$ subspace
\begin{gather}
	\hat c^{\dagger}_{\odot, j} = \frac{1}{\sqrt L} \sum_{n=0}^{L-1} \mathrm{e}^{-\mathrm{i} k_n j} \hat c^\dagger_{\odot, k_n}\; ,
	\quad
	\hat c^{\nodagger}_{\odot, j} = \frac{1}{\sqrt L} \sum_{n=0}^{L-1} \mathrm{e}^{\mathrm{i} k_n j} \hat c^\nodagger_{\odot, k_n}\; , \\
	\begin{aligned}
		\Rightarrow
		\hat \Pi_\odot\mathrm{e}^{\mathrm{i}\pi \hat n_{\odot, j}} \hat \Pi_\odot
		&=
		\hat \Pi_\odot \left( 1-2\hat n_{\odot, j} \right) \hat \Pi_\odot
		=
		1- \hat \Pi_\odot \left( \frac{2}{L} \sum_{n,m=0}^{L-1} \mathrm{e}^{-\mathrm{i}(k_n-k_m)j} \hat c^\dagger_{\odot, k_n} \hat c^\nodagger_{\odot, k_m} \right) \hat \Pi_\odot \\
		&=
		1 - \frac{2}{L} \hat c^\dagger_{\odot, k_n=0} \hat c^\nodagger_{\odot, k_n=0}
		\equiv
		1 - \frac{2}{L} \hat n_{\odot, k_n=0} \; .
	\end{aligned}
\end{gather}
In a similar manner, we expand the second sum in terms of rotations of both the outer and inner ring yielding
\begin{align}
	\hat H \ket{n_+,n_-} \ket{\mathbf k_N}
	&=
	\left(
		\underbrace{t \sum_{j=0}^{L-1}\left( \hat c^\dagger_j \hat c^\nodagger_{j+1} + \mathrm{h.c.} \right)}_{\sum\limits_{n=0}^{L-1} \varepsilon_{k_n} \hat c^\dagger_{k_n} \hat c^\nodagger_{k_n}} \left( 1 - \frac{2}{L} \hat n_{\odot, k=0} \right)
		-
		\left( \tilde s_j \hat c^\dagger_{k=k_0} \hat c^\nodagger_{\odot, k=0} + \mathrm{h.c.} \right)
	\right) \ket{n_+,n_-} \ket{\mathbf k_N} \; .
	\label{eq:supp:hubb-wheel:proj-mp-representation}
\end{align}
Using $\hat c^\dagger_{k=k_0} \hat c^\nodagger_{\odot, k=0} \ket{0,0}\ket{\mathbf k_N} = 0$ as well as $\left( 1 - \frac{2}{L} \hat n_{\odot, k=0}\right)\ket{0,0}\ket{\mathbf k_N} = \ket{0,0}\ket{\mathbf k_N}$, we immediately find that Slater determinants with empty $k=k_0$ modes are eigenstates of $\hat H$ with eigenvalues $E(\mathbf k_N) = \sum\limits_{l=1}^N \varepsilon_{k_l}$.
Having found~\cref{eq:supp:hubb-wheel:proj-mp-representation} the solution strategy for the many\hyp particle problem is straightforward.
Employing total particle\hyp number conservation, we decompose the many\hyp particle Hilbert space into orthogonal subspaces $\mathcal H_N$ with fixed total particle number $N$.
Each $N$\hyp particle subspace is then stratified into $4$\hyp dimensional subspaces $\mathcal H_{k^\prime, k^\pprime, \mathbf k_{N-2}}$ that are parametrized by a set of $N$ different modes $\mathbf k_N = \mathbf k_{N-2} \cup \left\{k^\prime, k^\pprime \right\}$ with $k_l, k^\prime, k^\pprime \neq k_0$, and we specified two distinct modes $k^\prime, k^\pprime$.
A subspace $\mathcal H_{k^\prime, k^\pprime, \mathbf k_{N-2}}$ is given by the linear hull of states
\begin{align}
	\ket{\mathbf k_{N-2}, k^\prime, k^\pprime}
	&=
	\alpha_0 \ket{\bm k_{N-2}, k^\prime, k^\pprime}
	+
	\alpha_{1_+} \ket{1,0}\ket{\bm k_{N-2}, k^\prime}
	+
	\alpha_{1_-} \ket{0,1} \ket{\bm k_{N-2}, k^\prime}
	+
	\alpha_2 \ket{1,1} \ket{\bm k_{N-2}} \; ,
	\label{eq:supp:mp-basis-states}
\end{align}
with complex coefficients $\alpha_{0,1_+,1_-,2} \in \mathbb C$.
Using orthogonality of the Slater determinants, it can be checked that this stratification yields a complete decomposition of the $N$\hyp particle Hilbert space by counting the dimensionalities.
Varying the set of modes $\mathbf k_{N-2} \cup \left\{k^\prime, k^\pprime \right\}$, the number of orthogonal basis states is obtained from the right hand side of~\cref{eq:supp:mp-basis-states} via
\begin{align}
	\operatorname{dim} \bigoplus_{\mathbf k_N, k^\prime, k^\pprime} \mathcal H_{k^\prime, k^\pprime, \mathbf k_{N-2}}
	&=
	\binom{L-1}{N} + 2\binom{L-1}{N-1} + \binom{L-1}{N-2}
	=
	\binom{L}{N} + \binom{L}{N-1}
	=
	\binom{L+1}{N}
	=
	\operatorname{dim} \mathcal H_N\; .
\end{align}
The last equality shows that indeed, the choosen parametrization generates a complete basis set for $\mathcal H_N$.
It is also easy to see that $\hat \Pi_\odot \hat H_\mathrm{lad} \hat \Pi_\odot$ does not mix states belonging to different subspaces $\mathcal H_{k^\prime, k^\pprime, \mathbf k_{N-2}}$ by noting that~\cref{eq:supp:hubb-wheel:proj-mp-representation} can only change the occupation of the $k=k_0$ modes.
Thus, we can solve the eigenvalue equation in each subspace separately, i.e., diagonalizing a $4\times 4$ matrix where, in the following, we abbreviate the Slater determinants suppressing the chosen set of $k$-values:
\begin{align}
	\ket{0,0}\ket{\mathbf k_{N-2}, k^\prime, k^\pprime} &\equiv \ket{0,0}, \;
	\ket{1,0}\ket{\mathbf k_{N-2}, k^\prime} \equiv \ket{1,0}, \; 
	\ket{0,1}\ket{\mathbf k_{N-2}, k^\prime} \equiv \ket{0,1} \; \text{and}\;
	\ket{1,1}\ket{\mathbf k_{N-2}} \equiv \ket{1,1} \; .
\end{align}
Using~\cref{eq:supp:hubb-wheel:proj-mp-representation} we immediately find
\begin{align}
	h_0 = \braket{n_+,n_- | \hat H | 0,0} &= \delta_{n_+,0} \delta_{n_-,0} E(\mathbf k_N)
	\quad \text{and} \quad
	h_2 = \braket{n_+,n_- | \hat H | 1,1}  = \delta_{n_+,1} \delta_{n_-,1} \left(E(\mathbf k_{N-2}) + \epsilon_{k_0} \right)\left( 1-\frac{2}{L}\right) \; ,
\end{align}
i.e., Slater determinants with empty or doubly occupied $k=k_0$ mode are both eigenstates of $\hat H$.
The remaining matrix elements evaluate to
\begin{equation}
	\begin{aligned}
		\braket{1,0 | \hat H | 1,0} &= \frac{\Delta_+\left( a \Delta_+ - 2\tilde s\right) + b}{1+\Delta_+^2} \; , \\
		\braket{0,1 | \hat H | 0,1} &= \frac{\Delta_-\left( a \Delta_- - 2\tilde s\right) + b}{1+\Delta_-^2} \; ,\\
		\braket{1,0 | \hat H | 0,1} &= \frac{a\Delta_+\Delta_- + \tilde s\left(\Delta_+ + \Delta_- \right) + b}{\sqrt{\left(1+\Delta_+^2\right)\left(1+\Delta_-^2\right)}} = \braket{0,1 | \hat H | 1,0} \; ,
	\end{aligned}
\end{equation}
with the definitions $a = E(\mathbf k_{N-1})\left(1-\frac{2}{L}\right)$, $b = \epsilon_{k_0} + E(\mathbf k_{N-1})$ as well as $E(\mathbf k_{N-1}) = E(\mathbf k_N) - \epsilon_{k^\prime}$.
Diagonalizing the $2\times 2$ matrix and setting $t\equiv 1$ as unit of energy finally yields the desired eigenvalues
\begin{equation}
	\begin{aligned}
		E_0(\mathbf k_{N-2},k^\prime, k^\pprime) &= E(\mathbf k_N) \\
		E_{1_\pm} (\mathbf k_{N-2},k^\prime) &= E(\mathbf k_{N-1}) \left(1-\frac{1}{L} \right) + 1 \pm \sqrt{\left(\frac{E(\mathbf k_{N-1})}{L} + 1\right)^2 + \tilde s^2} \\
		E_2(\mathbf k_{N}) &= \left(E(\mathbf k_{N-2}) +2 \right) \left(1-\frac{2}{L} \right) \; .
	\end{aligned}
	\label{eq:supp:hubb-wheel:mp-eigenvalues}
\end{equation}
These calculations imply a clustering of the many\hyp particle eigenvalues as shown in Fig. 3 in the main text.
For the set of all allowed modes $\mathbf k_{N-2} \cup \left\{ k^\prime, k^\pprime \right\}$ excluding $k=k_0$, each of the $4$ energies consitutes a bulk of many\hyp particle eigenvalues to the total spectrum with a bandwidth $\sim E(\mathbf k_N)$.
Furthermore, for large system sizes $L \gg 1$, the level spacing in the bulk of the clustered many\hyp particle eigenvalues scales as $\sim \frac{2\pi}{L}$.
Importantly, the clustered eigenvalues behave differently, depending on the occupations of the $k=k_0$ modes.
For the case of empty or doubly occupied $k=k_0$ modes, up to corrections $\sim \frac{1}{L}$ the spectrum is given by the summed single\hyp particle energies ($E(\mathbf k_N)$ and $\left(E(\mathbf k_{N-2})+2 \right)$).
If, however the $k=k_0$ mode is singly occupied, we obtain a separation of the clustered many\hyp particle eigenvalues $\sim \pm \lvert \tilde s\rvert$.
We want to point out that in the latter case, the overall dependency of the many\hyp particle eigenvalues on the ring\hyp to\hyp center coupling closely resembles that of single\hyp particle dispersion relation of the underlying fermionic ladder Hamiltonian upon replacing $\epsilon_k \rightarrow E(\mathbf k_{N-1})$.
This is not by accident but a result of the extensively scaling coordination number of the central site, restoring the single\hyp particle character.
The previous discussion suggest the definition two energy gaps $\Delta_{1},\Delta_{2}$ characterizing the transition to the \gls{BEC} phase as indicated in Fig. 3 in the main text.
Note that as soon as $\Delta_{1}>0$ the ground state is characterized by singly occupied $k=k_0$ modes, i.e., an odd $k=k_0$ parity that is separated from the clustered many\hyp body states with even $k=k_0$ parity and therefore, empty or doubly occupied $k=k_0$ modes.
Thus, $\Delta_{1}>0$ indicates a condensation of bosons into the $k=k_0$ mode.
However, only if $\Delta_{2}>0$ there is a finite gap between the different parity symmetry sectors, otherwise the clusters with even and odd $k=k_0$ parity are overlapping.
Evaluating the gaps using~\cref{eq:supp:hubb-wheel:mp-eigenvalues} in the limit $L\gg 1$ and defining $\rho = N/L$ yields
\begin{equation}
	\begin{aligned}
		\Delta_{1}
		&=
		-\left(1+\frac{2\sin(\pi\rho)}{\pi} \right) + \sqrt{\left(1-\frac{2\sin(\pi\rho)}{\pi} \right)^2 + \tilde s^2} \\
		\Delta_{2}
		&=
		-4L\frac{\sin(\pi\rho)}{\pi} - \left(1-\frac{2\sin(\pi\rho)}{\pi} \right) + \sqrt{\left(1+\frac{2\sin(\pi\rho)}{\pi} \right)^2 + \tilde s^2}
	\end{aligned}
\end{equation}
and thus the critical ring\hyp to\hyp center hoppings are given by
\begin{equation}
	\begin{aligned}
		s_{\mathrm{c},1}
		&=
		\frac{2}{\sqrt L}\sqrt{\frac{2\sin(\pi \rho)}{\pi}}
		\quad\text{and}\quad
		s_{\mathrm{c},2}
		&=
		2\sqrt L\sqrt{\frac{\sin(\pi\rho)}{\pi}\left(\frac{4\sin(\pi\rho)}{\pi} + 2\frac{1-\frac{2\sin(\pi\rho)}{\pi}}{L} - \frac{2}{L^2} \right)} \; .
	\end{aligned}
\end{equation}
\section{\label{sec:supp:numerical-details}Numerical details}%
\subsection{General remarks}%
\begin{figure}[!h]%
	\tikzsetnextfilename{wheelAsChain}%
	\begin{tikzpicture}
		\def\nmax{10}%
		\def\phase{7}%
		\def\radius{4}%
		\def\offset{2}%
%
%
		\node[star, draw, inner sep = 0, minimum width = 0.5em, line width = 0.02em] (center) {};%
		\foreach \n in {1,...,\nmax}%
		{%
			\node[circle, draw, inner sep = 0, minimum width = 0.5em, line width = 0.02em] at ($({\radius*cos((\n-1)/\nmax*360+\phase)},{\radius*sin((\n-1)/\nmax*360+\phase)})+(center)$) (ring\n) {};%
			\draw[densely dashed, draw=red!70!black, line width = 0.02em] (center) to (ring\n);%
			\node at ($({0.7*\radius*cos((\n-1+0.4)/\nmax*360)},{0.7*\radius*sin((\n-1+0.4)/\nmax*360)})+(center)$) {$s$};%
		}%
		\pgfmathparse{\nmax-1}%
		\def\nmaxmm{\pgfmathresult}%
		\foreach \n in {1,...,\nmaxmm}%
		{%
			\pgfmathparse{int(\n+1)}%
			\edef\npp{\pgfmathresult}%
			\draw[densely dotted, line width = 0.05em] (ring\n) to (ring\npp);%
			\node at ($({1.025*\radius*cos((\n-1+0.5)/\nmax*360+\phase)},{1.025*\radius*sin((\n-1+0.5)/\nmax*360+\phase)})+(center)$) {$t$};%
		}%
		\draw[densely dotted, line width = 0.05em] (ring\nmax) to (ring1);%
		\node at ($({1.025*\radius*cos((-0.5)/\nmax*360+\phase)},{1.025*\radius*sin((-0.5)/\nmax*360+\phase)})+(center)$) {$t$};%
%
%
		\pgfmathparse{int(\nmax/2)}%
		\edef\nHalf{\pgfmathresult}%
		\node[star, draw, inner sep = 0, minimum width = 0.5em, line width = 0.02em] at ($(0,-\radius-\offset)+(center)$) (chaincenter) {};%
		\draw[->,dashed, draw=black!30, line width = 0.05em] (center) to (chaincenter);%
		\foreach \n in {1,...,\nmax}%
		{%
			\node[circle, draw, inner sep = 0, minimum width = 0.5em, line width = 0.02em] at ($({\radius*cos((\n-1)/\nmax*360+\phase)},-\radius-\offset)+(center)$) (chain\n) {};%
			\pgfmathparse{cos((\n-1)/\nmax*360+\phase)}%
			\edef\nXPos{\pgfmathresult}%
			\ifthenelse{\lengthtest{\nXPos pt < 0pt}}%
			{%
				\draw[densely dashed, draw=red!70!black, looseness=1.1, bend left = 40] (chain\n) edge (chaincenter);%
			}%
			{%
				\draw[densely dashed, draw=red!70!black, looseness=1.1, bend left = -40] (chain\n) edge (chaincenter);%
			}%
		}%
		\pgfmathparse{\nmax-1}%
		\def\nmaxmm{\pgfmathresult}%
		\foreach \n in {1,...,\nmaxmm}%
		{%
			\pgfmathparse{int(\n+1)}%
			\edef\npp{\pgfmathresult}%
			\ifthenelse{\n = \nHalf}%
			{%
				\draw[densely dotted, line width = 0.05em] (chain\n) to (chain\npp);%
			}%
			{%
				\draw[densely dotted, looseness=1, bend left = 30, line width = 0.05em] (chain\n) to (chain\npp);%
			}%
		}%
		\draw[densely dotted, line width = 0.05em] (chain\nmax) to (chain1);%
%
%
		\begin{scope}[on background layer]%
			\foreach \n in {1,...,\nmax}%
			{%
				\draw[->,dashed, draw=black!30, line width = 0.05em] (ring\n) to (chain\n);%
			}%
		\end{scope}%
	\end{tikzpicture}
	\caption%
	{%
		\label{fig:wheelAsChain}%
		Chosen projection of the wheel geometry (top) onto a chain (bottom). %
		This reduces the long\hyp range interaction from the first to the last site that is replaced by multiple next-nearest-neighbor interactions compared to a straight forward \gls{PBC} implementation. %
		The center site (star) of the wheel is placed in the middle of this chain -- again in order to reduce the long\hyp range interaction to a minimum. %
	}
\end{figure}
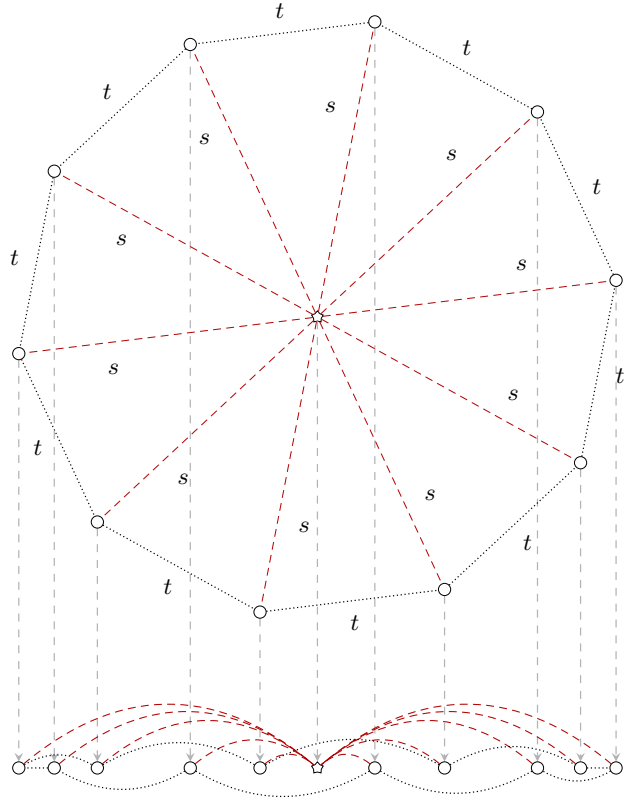%
All numerical results were obtained using the \gls{DMRG} \cite{White1992,White1993} in its \gls{MPS} representation \cite{Schollwock:2005p2117} implemented in the \symmps\ toolkit \cite{symmps}. %
More precisely, the calculations were performed with a bond dimension up to $1200$, which allowed the discarded weight to always stay below $2\cdot 10^{-8}$ and usually below $10^{-10}$. %
Since \gls{DMRG} works best in \gls{1D} systems, the wheel is projected onto a chain in a way that reduces the long\hyp range interaction to a (rather large) minimum, see \cref{fig:wheelAsChain}. %
\subsection{Observables}%
The observable of interest, as shown in~Fig. 4 in the main text, is the normalized condensate fraction of the distinguished $k_0$\hyp mode extrapolated to the thermodynamic limit. %
Note that in our calculations we chose $k_0=0$.
In order to obtain this quantity, we need to get the \gls{SPDM} -- in our case -- of the ground state, %
\begin{align}%
	\rho_{j, j^{\prime}}=\braket{\hat c^\dagger_{j^{\noprime}} \hat c^\nodagger_{j^{\prime}}} \;,%
\end{align}%
for multiple system sizes.
The condensate fraction is then obtained by Fourier transforming the \gls{SPDM}. %
\begin{align}%
	n_{k}= \frac{1}{L} \sum_{j,j^{\prime}} \mathrm{e}^{-2 i k (j - j^{\prime}) / L} \rho_{j, j^{\prime}}\;,%
\end{align}%
In order to be able to compare the condensate fractions for different system sizes, it is necessary to normalize them w.r.t the maximally possible value. %
This is given by~\cite{Tennie2017}
\begin{align}
	n_\mathrm{max}(L) = (L-N+1) \cdot N/L\;.
\end{align}
We chose four different sizes of the outer ring ($32$, $64$, $128$, $256$) and extrapolated these normalized results via a $\nicefrac1L$ fit.%

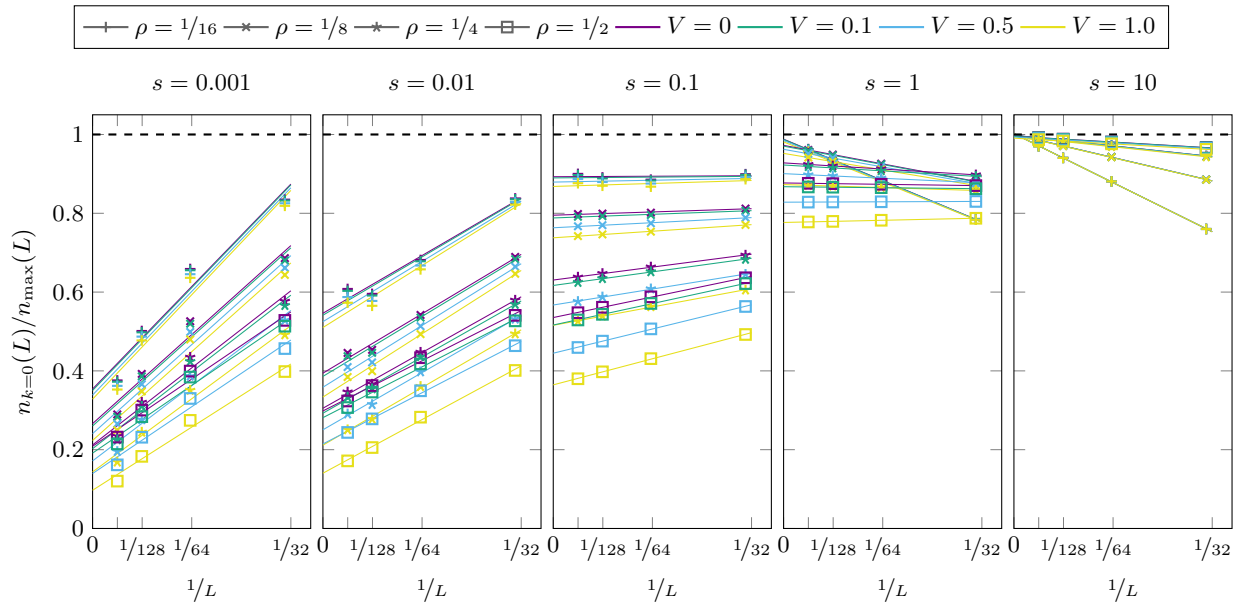
\begin{figure}[!h]%
	\centering%
	\tikzsetnextfilename{FiniteSizeScaling}%
	\begin{tikzpicture}%
		\begin{groupplot}%
		[%
			group style = %
			{%
				group size 			=	5 by 1,%
				horizontal sep		=	0.5em,%
				vertical sep		=	0em,%
				x descriptions at	=	edge bottom,%
				y descriptions at	=	edge left,%
			},%
			width				=	0.25\textwidth,%
			height				=	0.3\textheight,%
			xlabel				=	{$\nicefrac1L$},%
			ylabel				=	{$n_{k=0}(L)/n_\mathrm{max}(L)$},%
			legend style		=	{at = {(0.97,0.03)}, anchor = south east},%
			legend columns		=	8,%
			ymin				=	0,%
			ymax				=	1.05,%
			xmin				=	0,
			xtick				=	{0, 1.0/256.0, 1.0/128.0, 1.0/64.0, 1.0/32.0},
			xticklabels			=	{$0$, $$, $\nicefrac{1}{128}$, $\nicefrac{1}{64}$, $\nicefrac{1}{32}$},
			scaled x ticks		=	false,
		]%
			\def\Nfacs{0/2, 1/4, 2/8, 3/16}%
			\def\Vs{0/0, 1/0.1, 2/0.5, 3/1.0}%
			\nextgroupplot%
			[%
				title			=	{$s=0.001$},%
				title style		=	{name = title},%
				ylabel style	=	{name = ylabel},%
				legend to name	=	{CommonLegend},%
			]%
				\draw[dashed, thick, black] (axis cs:0,1) -- (axis cs:0.05,1);%
				\foreach \NfacIt/\Nfac in \Nfacs%
				{%
					\pgfmathsetmacro{\marker}{\gpmarkers[mod(\NfacIt,dim(\gpmarkers))]}%
					\pgfmathsetmacro{\n}{int(32.0/\Nfac)}%
					\edef\markerlegend%
					{%
						\noexpand\addlegendentry{$\rho=\noexpand\nicefrac{1}{\n}$};%
						\noexpand\addlegendimage{mark=\marker, color=black!60, thick};%
					}\markerlegend%
				}%
				\foreach \VIt/\V in \Vs%
				{%
					\pgfmathsetmacro{\cl}{\gpcolors[mod(\VIt,dim(\gpcolors))]}%
					\edef\colorlegend%
					{%
						\noexpand\addlegendentry{$V=\V$};%
						\noexpand\addlegendimage{color=\cl, thick};%
					}\colorlegend%
				}%
				\foreach \NfacIt/\Nfac in \Nfacs%
				{%
					\pgfmathsetmacro{\marker}{\gpmarkers[mod(\NfacIt,dim(\gpmarkers))]}%
					\foreach \VIt/\V in \Vs%
					{%
						\pgfmathsetmacro{\cl}{\gpcolors[mod(\VIt,dim(\gpcolors))]}%
						\edef\plot%
						{%
							\noexpand\addplot%
							[%
								no markers,%
								forget plot,%
							]%
								gnuplot%
								[%
									raw gnuplot%
								]%
									{%
										set fit quiet;%
										set print "../data/autogen/fft_k0_s_0.001_V_\V_Nfac_\Nfac.fit";%
										print "a aerror b berror";%
										set fit errorvariables;%
										b=1;%
										a=1;%
										s="0.001";%
										V="\V";%
										f(x) = a+b*x;%
										fit f(x) '<grep " -'.s.' '.V.' " ../data/fft_k0_vs_s.dat' u (1.0/$1):(($4*32.0/($1-1.0)==\Nfac)?$5/($4/$1*($1-$4+1.0)):0/0) via a,b;%
										print a,a_err,b,b_err;%
									};%
							\noexpand\pgfplotstableread[header=true]{../data/autogen/fft_k0_s_0.001_V_\V_Nfac_\Nfac.fit}{\noexpand\fit}%
							\noexpand\pgfplotstablegetelem{0}{a}\noexpand\of\noexpand\fit%
							\noexpand\edef\noexpand\fittedA{\noexpand\pgfplotsretval}%
							\noexpand\pgfplotstablegetelem{0}{aerror}\noexpand\of\noexpand\fit%
							\noexpand\edef\noexpand\fittedAError{\noexpand\pgfplotsretval}%
							\noexpand\pgfplotstablegetelem{0}{b}\noexpand\of\noexpand\fit%
							\noexpand\edef\noexpand\fittedB{\noexpand\pgfplotsretval}%
							\noexpand\pgfplotstablegetelem{0}{berror}\noexpand\of\noexpand\fit%
							\noexpand\edef\noexpand\fittedBError{\noexpand\pgfplotsretval}%
							\noexpand\addplot%
							[%
								domain 	= 	0:1.0/32,%
								color	=	\cl,%
								forget plot,%
							]%
							{linearFct(\noexpand\fittedB,\noexpand\fittedA)};%
						}\plot%
					}%
				}%
				\foreach \NfacIt/\Nfac in \Nfacs%
				{%
					\pgfmathsetmacro{\marker}{\gpmarkers[mod(\NfacIt,dim(\gpmarkers))]}%
					\foreach \VIt/\V in \Vs%
					{%
						\pgfmathsetmacro{\cl}{\gpcolors[mod(\VIt,dim(\gpcolors))]}%
						\edef\plot%
						{%
							\noexpand\addplot%
							[%
								mark		=	\marker,%
								color		=	\cl,%
								thick,%
								only marks,%
								forget plot,%
							]%
								table%
								[%
									x expr	= %
									{%
										\noexpand\thisrowno{1}==-0.001%
										?%
										(%
											\noexpand\thisrowno{2}==\V%
											?%
											(%
												round(\noexpand\thisrowno{3}*32.0/(1.0*\noexpand\thisrowno{0}-1.0))==1.0*\Nfac%
												?%
												1.0/\noexpand\thisrowno{0}%
												:%
												NaN%
											)%
											:%
											NaN%
										)%
										:%
										NaN%
									},%
									y expr	=	 \noexpand\thisrowno{4}/(\noexpand\thisrowno{3}*(\noexpand\thisrowno{0}-(\noexpand\thisrowno{3})+1.0)/(\noexpand\thisrowno{0})),%
								]%
							{../data/fft_k0_vs_s.dat};%
						}\plot%
					}%
				}%
			\nextgroupplot%
			[%
				title	=	{$s=0.01$},%
			]%
				\draw[dashed, thick, black] (axis cs:0,1) -- (axis cs:0.05,1);%
				\foreach \NfacIt/\Nfac in \Nfacs%
				{%
					\pgfmathsetmacro{\marker}{\gpmarkers[mod(\NfacIt,dim(\gpmarkers))]}%
					\foreach \VIt/\V in \Vs%
					{%
						\pgfmathsetmacro{\cl}{\gpcolors[mod(\VIt,dim(\gpcolors))]}%
						\edef\plot%
						{%
							\noexpand\addplot%
							[%
								no markers,%
								forget plot,%
							]%
								gnuplot%
								[%
									raw gnuplot%
								]%
									{%
										set fit quiet;%
										set print "../data/autogen/fft_k0_s_0.01_V_\V_Nfac_\Nfac.fit";%
										print "a aerror b berror";%
										set fit errorvariables;%
										b=1;%
										a=1;%
										s="0.01";%
										V="\V";%
										f(x) = a+b*x;%
										fit f(x) '<grep " -'.s.' '.V.' " ../data/fft_k0_vs_s.dat' u (1.0/$1):(($4*32.0/($1-1.0)==\Nfac)?$5/($4/$1*($1-$4+1.0)):0/0) via a,b;%
										print a,a_err,b,b_err;%
									};%
							\noexpand\pgfplotstableread[header=true]{../data/autogen/fft_k0_s_0.01_V_\V_Nfac_\Nfac.fit}{\noexpand\fit}%
							\noexpand\pgfplotstablegetelem{0}{a}\noexpand\of\noexpand\fit%
							\noexpand\edef\noexpand\fittedA{\noexpand\pgfplotsretval}%
							\noexpand\pgfplotstablegetelem{0}{aerror}\noexpand\of\noexpand\fit%
							\noexpand\edef\noexpand\fittedAError{\noexpand\pgfplotsretval}%
							\noexpand\pgfplotstablegetelem{0}{b}\noexpand\of\noexpand\fit%
							\noexpand\edef\noexpand\fittedB{\noexpand\pgfplotsretval}%
							\noexpand\pgfplotstablegetelem{0}{berror}\noexpand\of\noexpand\fit%
							\noexpand\edef\noexpand\fittedBError{\noexpand\pgfplotsretval}%
							\noexpand\addplot%
							[%
								domain 	= 	0:1.0/32,%
								color	=	\cl,%
								forget plot,%
							]%
							{linearFct(\noexpand\fittedB,\noexpand\fittedA)};%
						}\plot%
					}%
				}%
				\foreach \NfacIt/\Nfac in \Nfacs%
				{%
					\pgfmathsetmacro{\marker}{\gpmarkers[mod(\NfacIt,dim(\gpmarkers))]}%
					\foreach \VIt/\V in \Vs%
					{%
						\pgfmathsetmacro{\cl}{\gpcolors[mod(\VIt,dim(\gpcolors))]}%
						\edef\plot%
						{%
							\noexpand\addplot%
							[%
								mark	=	\marker,%
								color	=	\cl,%
								thick,%
								only marks,%
								forget plot,%
							]%
								table%
								[%
									x expr	= %
									{%
										\noexpand\thisrowno{1}==-0.01%
										?%
										(%
											\noexpand\thisrowno{2}==\V%
											?%
											(%
												round(\noexpand\thisrowno{3}*32.0/(1.0*\noexpand\thisrowno{0}-1.0))==1.0*\Nfac%
												?%
												1.0/\noexpand\thisrowno{0}%
												:%
												NaN%
											)%
											:%
											NaN%
										)%
										:%
										NaN%
									},%
									y expr	= \noexpand\thisrowno{4}/(\noexpand\thisrowno{3}*(\noexpand\thisrowno{0}-(\noexpand\thisrowno{3})+1.0)/(\noexpand\thisrowno{0})),%
							]%
							{../data/fft_k0_vs_s.dat};%
						}\plot%
					}%
				}%
			\nextgroupplot%
			[%
				title	=	{$s=0.1$},%
			]%
				\draw[dashed, thick, black] (axis cs:0,1) -- (axis cs:0.05,1);%
				\foreach \NfacIt/\Nfac in \Nfacs%
				{%
					\pgfmathsetmacro{\marker}{\gpmarkers[mod(\NfacIt,dim(\gpmarkers))]}%
					\foreach \VIt/\V in \Vs%
					{%
						\pgfmathsetmacro{\cl}{\gpcolors[mod(\VIt,dim(\gpcolors))]}%
						\edef\plot%
						{%
							\noexpand\addplot%
							[%
								no markers,%
								forget plot,%
							]%
								gnuplot%
								[%
									raw gnuplot%
								]%
									{%
										set fit quiet;%
										set print "../data/autogen/fft_k0_s_0.1_V_\V_Nfac_\Nfac.fit";%
										print "a aerror b berror";%
										set fit errorvariables;%
										b=1;%
										a=1;%
										s="0.1";%
										V="\V";%
										f(x) = a+b*x;%
										fit f(x) '<grep " -'.s.' '.V.' " ../data/fft_k0_vs_s.dat' u (1.0/$1):(($4*32.0/($1-1.0)==\Nfac)?$5/($4/$1*($1-$4+1.0)):0/0) via a,b;%
										print a,a_err,b,b_err;%
									};%
							\noexpand\pgfplotstableread[header=true]{../data/autogen/fft_k0_s_0.1_V_\V_Nfac_\Nfac.fit}{\noexpand\fit}%
							\noexpand\pgfplotstablegetelem{0}{a}\noexpand\of\noexpand\fit%
							\noexpand\edef\noexpand\fittedA{\noexpand\pgfplotsretval}%
							\noexpand\pgfplotstablegetelem{0}{aerror}\noexpand\of\noexpand\fit%
							\noexpand\edef\noexpand\fittedAError{\noexpand\pgfplotsretval}%
							\noexpand\pgfplotstablegetelem{0}{b}\noexpand\of\noexpand\fit%
							\noexpand\edef\noexpand\fittedB{\noexpand\pgfplotsretval}%
							\noexpand\pgfplotstablegetelem{0}{berror}\noexpand\of\noexpand\fit%
							\noexpand\edef\noexpand\fittedBError{\noexpand\pgfplotsretval}%
							\noexpand\addplot%
							[%
								domain 	= 	0:1.0/32,%
								color	=	\cl,%
								forget plot,%
							]%
							{linearFct(\noexpand\fittedB,\noexpand\fittedA)};%
						}\plot%
					}%
				}%
				\foreach \NfacIt/\Nfac in \Nfacs%
				{%
					\pgfmathsetmacro{\marker}{\gpmarkers[mod(\NfacIt,dim(\gpmarkers))]}%
					\foreach \VIt/\V in \Vs%
					{%
						\pgfmathsetmacro{\cl}{\gpcolors[mod(\VIt,dim(\gpcolors))]}%
						\edef\plot%
						{%
							\noexpand\addplot%
							[%
								mark	=	\marker,%
								color	=	\cl,%
								thick,%
								only marks,%
								forget plot,%
							]%
								table%
								[%
									x expr	= %
									{%
										\noexpand\thisrowno{1}==-0.1%
										?%
										(%
											\noexpand\thisrowno{2}==\V%
											?%
											(%
												round(\noexpand\thisrowno{3}*32.0/(1.0*\noexpand\thisrowno{0}-1.0))==1.0*\Nfac%
												?%
												1.0/\noexpand\thisrowno{0}%
												:%
												NaN%
											)%
											:%
											NaN%
										)%
										:%
										NaN%
									},%
									y expr	= \noexpand\thisrowno{4} / (\noexpand\thisrowno{3}*(\noexpand\thisrowno{0}-(\noexpand\thisrowno{3})+1.0)/(\noexpand\thisrowno{0})),%
								]%
							{../data/fft_k0_vs_s.dat};%
						}\plot%
					}%
				}%
			\nextgroupplot%
			[%
				title				=	{$s=1$},%
			]%
				\draw[dashed, thick, black] (axis cs:0,1) -- (axis cs:0.05,1);%
				\foreach \NfacIt/\Nfac in \Nfacs%
				{%
					\pgfmathsetmacro{\marker}{\gpmarkers[mod(\NfacIt,dim(\gpmarkers))]}%
					\foreach \VIt/\V in \Vs%
					{%
						\pgfmathsetmacro{\cl}{\gpcolors[mod(\VIt,dim(\gpcolors))]}%
						\edef\plot%
						{%
							\noexpand\addplot%
							[%
								no markers,%
								forget plot,%
							]%
								gnuplot%
								[%
									raw gnuplot%
								]%
									{%
										set fit quiet;%
										set print "../data/autogen/fft_k0_s_1_V_\V_Nfac_\Nfac.fit";%
										print "a aerror b berror";%
										set fit errorvariables;%
										b=1;%
										a=1;%
										s="1";%
										V="\V";%
										f(x) = a+b*x;%
										fit f(x) '<grep " -'.s.' '.V.' " ../data/fft_k0_vs_s.dat' u (1.0/$1):(($4*32.0/($1-1.0)==\Nfac)?$5/($4/$1*($1-$4+1.0)):0/0) via a,b;%
										print a,a_err,b,b_err;%
									};%
							\noexpand\pgfplotstableread[header=true]{../data/autogen/fft_k0_s_1_V_\V_Nfac_\Nfac.fit}{\noexpand\fit}%
							\noexpand\pgfplotstablegetelem{0}{a}\noexpand\of\noexpand\fit%
							\noexpand\edef\noexpand\fittedA{\noexpand\pgfplotsretval}%
							\noexpand\pgfplotstablegetelem{0}{aerror}\noexpand\of\noexpand\fit%
							\noexpand\edef\noexpand\fittedAError{\noexpand\pgfplotsretval}%
							\noexpand\pgfplotstablegetelem{0}{b}\noexpand\of\noexpand\fit%
							\noexpand\edef\noexpand\fittedB{\noexpand\pgfplotsretval}%
							\noexpand\pgfplotstablegetelem{0}{berror}\noexpand\of\noexpand\fit%
							\noexpand\edef\noexpand\fittedBError{\noexpand\pgfplotsretval}%
							\noexpand\addplot%
							[%
								domain 	= 	0:1.0/32,%
								color	=	\cl,%
								forget plot,%
							]%
							{linearFct(\noexpand\fittedB,\noexpand\fittedA)};%
						}\plot%
					}%
				}%
				\foreach \NfacIt/\Nfac in \Nfacs%
				{%
					\pgfmathsetmacro{\marker}{\gpmarkers[mod(\NfacIt,dim(\gpmarkers))]}%
					\foreach \VIt/\V in \Vs%
					{%
						\pgfmathsetmacro{\cl}{\gpcolors[mod(\VIt,dim(\gpcolors))]}%
						\edef\plot%
						{%
							\noexpand\addplot%
							[%
								mark	=	\marker,%
								color	=	\cl,%
								thick,%
								only marks,%
								forget plot,%
							]%
								table%
								[%
									x expr	= %
									{%
										\noexpand\thisrowno{1}==-1%
										?%
										(%
											\noexpand\thisrowno{2}==\V%
											?%
											(%
												round(\noexpand\thisrowno{3}*32.0/(1.0*\noexpand\thisrowno{0}-1.0))==1.0*\Nfac%
												?%
												1.0/\noexpand\thisrowno{0}%
												:%
												NaN%
											)%
											:%
											NaN%
										)%
										:%
										NaN%
									},%
									y expr	= \noexpand\thisrowno{4} / (\noexpand\thisrowno{3}*(\noexpand\thisrowno{0}-(\noexpand\thisrowno{3})+1.0)/(\noexpand\thisrowno{0})),%
								]%
							{../data/fft_k0_vs_s.dat};%
						}\plot%
					}%
				}%
			\nextgroupplot%
			[%
				title				=	{$s=10$},%
				xticklabel style	=	{name=xlabel},%
			]%
				\foreach \NfacIt/\Nfac in \Nfacs%
				{%
					\pgfmathsetmacro{\marker}{\gpmarkers[mod(\NfacIt,dim(\gpmarkers))]}%
					\foreach \VIt/\V in \Vs%
					{%
						\pgfmathsetmacro{\cl}{\gpcolors[mod(\VIt,dim(\gpcolors))]}%
						\edef\plot%
						{%
							\noexpand\addplot%
							[%
								no markers,%
								forget plot,%
							]%
								gnuplot%
								[%
									raw gnuplot%
								]%
									{%
										set fit quiet;%
										set print "../data/autogen/fft_k0_s_10_V_\V_Nfac_\Nfac.fit";%
										print "a aerror b berror";%
										set fit errorvariables;%
										b=1;%
										a=1;%
										s="10";%
										V="\V";%
										f(x) = a+b*x;%
										fit f(x) '<grep " -'.s.' '.V.' " ../data/fft_k0_vs_s.dat' u (1.0/$1):(($4*32.0/($1-1.0)==\Nfac)?$5/($4/$1*($1-$4+1.0)):0/0) via a,b;%
										print a,a_err,b,b_err;%
									};%
							\noexpand\pgfplotstableread[header=true]{../data/autogen/fft_k0_s_10_V_\V_Nfac_\Nfac.fit}{\noexpand\fit}%
							\noexpand\pgfplotstablegetelem{0}{a}\noexpand\of\noexpand\fit%
							\noexpand\edef\noexpand\fittedA{\noexpand\pgfplotsretval}%
							\noexpand\pgfplotstablegetelem{0}{aerror}\noexpand\of\noexpand\fit%
							\noexpand\edef\noexpand\fittedAError{\noexpand\pgfplotsretval}%
							\noexpand\pgfplotstablegetelem{0}{b}\noexpand\of\noexpand\fit%
							\noexpand\edef\noexpand\fittedB{\noexpand\pgfplotsretval}%
							\noexpand\pgfplotstablegetelem{0}{berror}\noexpand\of\noexpand\fit%
							\noexpand\edef\noexpand\fittedBError{\noexpand\pgfplotsretval}%
							\noexpand\addplot%
							[%
								domain 	= 	0:1.0/32,%
								color	=	\cl,%
								forget plot,%
							]%
							{linearFct(\noexpand\fittedB,\noexpand\fittedA)};%
						}\plot%
					}%
				}%
				\draw[dashed, thick, black] (axis cs:0,1) -- (axis cs:0.05,1);%
				\foreach \NfacIt/\Nfac in \Nfacs%
				{%
					\pgfmathsetmacro{\marker}{\gpmarkers[mod(\NfacIt,dim(\gpmarkers))]}%
					\foreach \VIt/\V in \Vs%
					{%
						\pgfmathsetmacro{\cl}{\gpcolors[mod(\VIt,dim(\gpcolors))]}%
						\edef\plot%
						{%
							\noexpand\addplot%
							[%
								mark	=	\marker,%
								color	=	\cl,%
								thick,%
								only marks,%
								forget plot,%
							]%
								table%
								[%
									x expr	= %
									{%
										\noexpand\thisrowno{1}==-10%
										?%
										(%
											\noexpand\thisrowno{2}==\V%
											?%
											(%
												round(\noexpand\thisrowno{3}*32.0/(1.0*\noexpand\thisrowno{0}-1.0))==1.0*\Nfac%
												?%
												1.0/\noexpand\thisrowno{0}%
												:%
												NaN%
											)%
											:%
											NaN%
										)%
										:%
										NaN%
									},%
									y expr	= \noexpand\thisrowno{4} / (\noexpand\thisrowno{3}*(\noexpand\thisrowno{0}-(\noexpand\thisrowno{3})+1.0)/(\noexpand\thisrowno{0})),%
								]%
							{../data/fft_k0_vs_s.dat};%
						}\plot%
					}%
				}%
		\end{groupplot}%
		\node[fit = (title) (ylabel) (xlabel)] (all) {};%
		\node[anchor=south, at = (all.north)] {\pgfplotslegendfromname{CommonLegend}};%
	\end{tikzpicture}%
	\caption%
	{%
		\label{fig:condensate-fraction:Finite-size-fit}%
		Examples for the finite size scaling.
		For reasonably large densities, i.e., dense many\hyp particle eigenstate clusters, we observe excellent $\nicefrac{1}{L}$ scaling to finite condensate fractions (normalized to the maximally possible condensate density $n_\mathrm{max}(L)$.
		In the highly dilute limite, the many\hyp particle spectrum is very sparse, causing an abrupt opening of the gap $\Delta_2$, which also effectively suppresses scatterings caused by the interactions.
	}%
\end{figure}
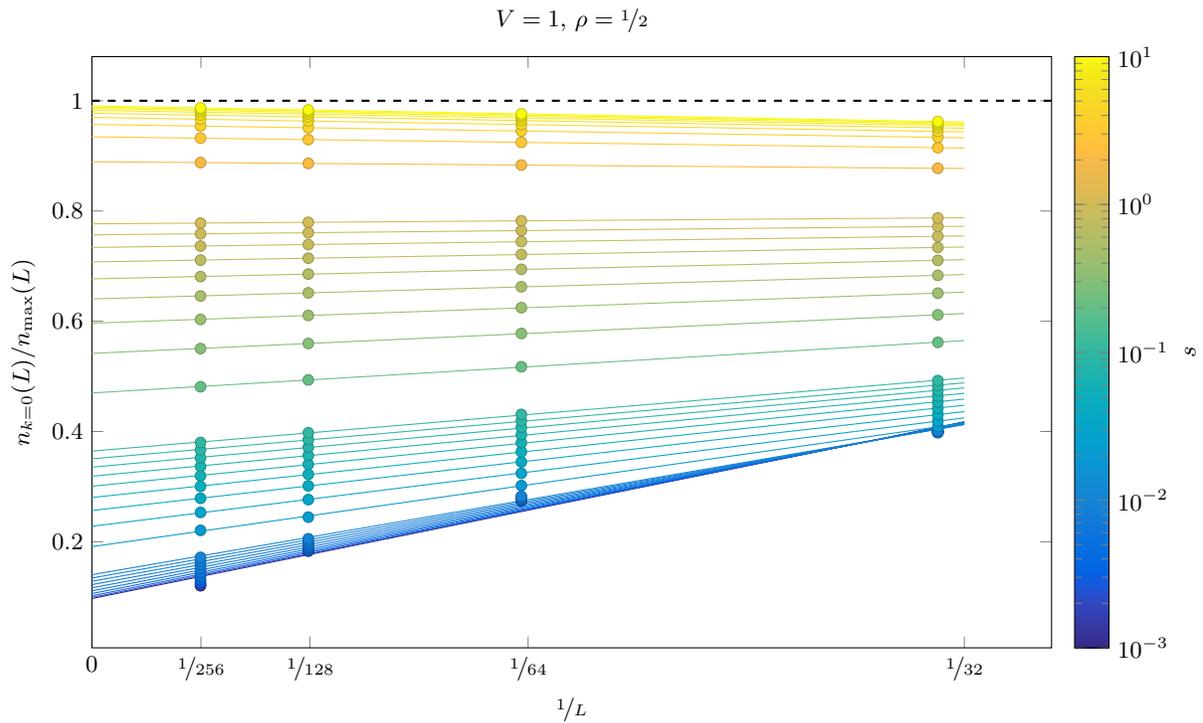
\begin{figure}[!h]%
	\centering%
	\tikzsetnextfilename{FiniteSizeScalingSingleVSingleMu}%
	\def\pmmin{0.001}
	\def\pmmax{10}
	\begin{tikzpicture}%
		\begin{axis}%
		[%
			title				=	{$V=1$, $\rho=\nicefrac12$},%
			width				=	0.8\textwidth,%
			height				=	0.4\textheight,%
			xlabel				=	{$\nicefrac1L$},%
			ylabel				=	{$n_{k=0}(L)/n_\mathrm{max}(L)$},%
			xmin				=	0,
			xtick				=	{0, 1.0/256.0, 1.0/128.0, 1.0/64.0, 1.0/32.0},
			xticklabels			=	{$0$, $\nicefrac{1}{256}$, $\nicefrac{1}{128}$, $\nicefrac{1}{64}$, $\nicefrac{1}{32}$},
			colorbar,
			point meta min	= \pmmin,
			point meta max	= \pmmax,
			every colorbar/.append style =
			{
				ymode	=	log,
				width	= {13pt},
				ylabel	= {$s$},
			},
			scaled x ticks		=	false,
		]%
			\def\Nfac{16}%
			\def\V{1.0}%
			\def\Ss{0.001, 0.002, 0.003, 0.004, 0.005, 0.006, 0.007, 0.008, 0.009, 0.01, 0.02, 0.03, 0.04, 0.05, 0.06, 0.07, 0.08, 0.09, 0.1, 0.2, 0.3, 0.4, 0.5, 0.6, 0.7, 0.8, 0.9, 1, 2, 3, 4, 5, 6, 7, 8, 9, 10}%
			\draw[dashed, thick, black] (axis cs:0,1) -- (axis cs:0.05,1);%

			\foreach \s in \Ss%
			{
				\pgfmathsetmacro{\pome}{(ln(\s)/ln(10)-ln(\pmmin)/ln(10))/(ln(\pmmax)/ln(10)-ln(\pmmin)/ln(10))*(\pmmax-\pmmin)+\pmmin}
				\edef\plot%
				{%
					\noexpand\addplot
					[
						no markers, 
						forget plot,
					] 
						gnuplot 
						[
							raw gnuplot
						] 
							{
								set fit quiet;
								set print "../data/autogen/fft_k0_s_\s.fit";
								print "a aerror b berror";
								set fit errorvariables;
								b=1;
								a=1;
								s="\s";
								f(x) = a+b*x;
								fit f(x) '<grep " -'.s.' \V " ../data/fft_k0_vs_s.dat' u (1.0/$1):(($4*32.0/($1-1.0)==\Nfac)?$5/($4/$1*($1-$4+1.0)):0/0) via a,b;
								print a,a_err,b,b_err;
							};

					\noexpand\pgfplotstableread[header=true]{../data/autogen/fft_k0_s_\s.fit}{\noexpand\fit}
					\noexpand\pgfplotstablegetelem{0}{a}\noexpand\of\noexpand\fit 
					\noexpand\edef\noexpand\fittedA{\noexpand\pgfplotsretval}
					\noexpand\pgfplotstablegetelem{0}{aerror}\noexpand\of\noexpand\fit 
					\noexpand\edef\noexpand\fittedAError{\noexpand\pgfplotsretval}
					\noexpand\pgfplotstablegetelem{0}{b}\noexpand\of\noexpand\fit 
					\noexpand\edef\noexpand\fittedB{\noexpand\pgfplotsretval}
					\noexpand\pgfplotstablegetelem{0}{berror}\noexpand\of\noexpand\fit 
					\noexpand\edef\noexpand\fittedBError{\noexpand\pgfplotsretval}
					\noexpand\addplot
					[
						domain = 0:1.0/32,
						point meta = \pome,
						mesh,
						forget plot,%
					]
					{linearFct(\noexpand\fittedB,\noexpand\fittedA)};
				}\plot
			}

			\foreach \s in \Ss%
			{
				\pgfmathsetmacro{\pome}{(ln(\s)/ln(10)-ln(\pmmin)/ln(10))/(ln(\pmmax)/ln(10)-ln(\pmmin)/ln(10))*(\pmmax-\pmmin)+\pmmin}
				\edef\plot%
				{%
					\noexpand\addplot%
					[%
						point meta	=	\pome,
						only marks,
						scatter,
						forget plot,%
					]%
						table%
						[%
							x expr	= %
							{%
								\noexpand\thisrowno{1}==-\s%
								?%
								(%
									\noexpand\thisrowno{2}==\V%
									?%
									(%
										round(\noexpand\thisrowno{3}*32.0/(1.0*\noexpand\thisrowno{0}-1.0))==1.0*\Nfac%
										?%
										1.0/\noexpand\thisrowno{0}%
										:%
										NaN%
									)%
									:%
									NaN%
								)%
								:%
								NaN%
							},%
							y expr	=	 \noexpand\thisrowno{4}/(\noexpand\thisrowno{3}*(\noexpand\thisrowno{0}-(\noexpand\thisrowno{3})+1.0)/(\noexpand\thisrowno{0})),%
						]%
						{../data/fft_k0_vs_s.dat};%
				}\plot%
			}%
		\end{axis}%
	\end{tikzpicture}%
	\caption%
	{%
		\label{fig:condensate-fraction:Finite-size-fit-singleVrho-multipleS}%
		Examples for the finite size scaling.
		For density $\rho=\nicefrac12$, we observe excellent $\nicefrac{1}{L}$ scaling to finite condensate fractions (normalized to the maximally possible condensate density $n_\mathrm{max}(L)$) for all ring-to-center hoppings larger than $s > 10^{-2}$.
	}%
\end{figure}

%
In~\cref{fig:condensate-fraction:Finite-size-fit}, a subset of the used data is shown, namely all data for $s=0.001, 0.01, 0.1, 1, 10$. %
The chosen $\nicefrac1L$ fit is only valid if the density is large enough. %
This can be addressed to the fact that in the highly diluted regime the many\hyp particle spectrum becomes very sparse and causes an abrupt opening of the gap $\Delta_2$. %
Also note that this sparseness can effectively suppresses the effect of interactions. %
Furthermore, the difference of the normalized condensate fraction w.r.t. the interaction strength becomes more pronounced the more particles can interact with each other. %
If the ring\hyp to\hyp center hopping becomes to small ($s < 10^{-2}$) additional effects come into play and interfere with the $\nicefrac1L$ fit.
This can also be seen in \cref{fig:condensate-fraction:Finite-size-fit-singleVrho-multipleS}, in which the fits for the highest density ($\rho=\nicefrac12$) and the highest interaction strength ($V=1$) are shown.
\end{document}